\documentclass[english,letterpaper,floatfix,aps,prx,amsmath,amssymb,twocolumn,superscriptaddress,showpacs]{revtex4-1}
\usepackage[T1]{fontenc}
\usepackage[latin9]{inputenc}
\usepackage{amsmath}
\usepackage{amssymb}
\usepackage{graphicx}
\usepackage{esint}
\usepackage{hyperref}

\newcommand{\appsinglelevelQD}{S1.1 in~\cite{Supplemental_Material}}
\newcommand{\appdoublelevelQD}{S1.2 in~\cite{Supplemental_Material}}
\newcommand{\appone}{S2 in~\cite{Supplemental_Material}}
\newcommand{\apponeBsymmetries}{S2.2 in~\cite{Supplemental_Material}}
\newcommand{\apponeChermitianmatrices}{S2.3 in~\cite{Supplemental_Material}}
\newcommand{\apponeDZfromgeometricphase}{S2.4 in~\cite{Supplemental_Material}}
\newcommand{\apptwotopJJ}{S3 in~\cite{Supplemental_Material}}
\newcommand{\appthreegeometricfromwaiting}{S4 in~\cite{Supplemental_Material}}
\newcommand{\appthreegeometricfromwaitingshort}{S4}

\makeatletter
\usepackage{verbatim}

\makeatother

\usepackage{babel}
\begin{document}

\title{Fractional charges in conventional sequential electron tunneling}

\author{Roman-Pascal Riwar}

\affiliation{JARA Institute for Quantum Information (PGI-11), Forschungszentrum
J\"ulich, 52425 J\"ulich, Germany}

\begin{abstract}
The notion of fractional charges was up until now reserved for quasiparticle excitations emerging in strongly correlated quantum systems, such as Laughlin states in the fractional quantum Hall effect, Luttinger quasiparticles, or parafermions. Here, we consider topological transitions in the full counting statistics of standard sequential electron tunneling, and find that they lead to the same type of charge fractionalization - strikingly without requiring exotic quantum correlations. This conclusion relies on the realization that fundamental integer charge quantization fixes the global properties of the transport statistics whereas fractional charges can only be well-defined locally. We then show that the reconciliation of these two contradicting notions results in a nontrivially quantized geometric phase defined in the detector space. In doing so, we show that detector degrees of freedom can be used to describe topological transitions in nonequilibrium open quantum systems. Moreover, the quantized geometric phase reveals a profound analogy between the fractional charge effect in sequential tunneling and fractional Josephson effect in topological superconducting junctions, where likewise the Majorana- or parafermions exhibit a charge which is at odds with the Cooper pair charge as the underlying unit of the supercurrent. In order to provide means for an experimental verification of our claims, we demonstrate the fractional nature of transport statistics at the example of highly feasible transport models, such as weakly tunnel-coupled quantum dots or charge islands. We then show that the geometric phase can be accessed through the detector's waiting time distribution. Finally, we find that topological transitions in the transport statistics could even lead to new applications, such as the unexpected possibility to directly measure features beyond the resolution limit of a detector.
\end{abstract}

\pacs{05.40.-a, 06.90.+v, 64.70.-p, 73.23.Hk}
\maketitle

\section{Introduction}

The fundamental unit, in which transport is exchanged between electronic systems, is the elementary charge of the electron, $e$. However, in the solid state, strong quantum correlations and topological phase transitions are known to
lead to the emergence of particles and excitations which appear to carry, due
to their collective nature, a charge different from $e$. The probably best known example is the doubly
charged Cooper pair in superconductors. Notably, there can also appear excitations
with a \textit{fractional} charge, such as Laughlin quasiparticles
in the fractional quantum Hall effect \cite{Laughlin_1983}, electronic
realizations of parafermions \cite{Alicea_2016}, and quasiparticles
in Luttinger liquid nanowires \cite{Pham_2000,Trauzettel_2004,Steinberg_2007}. 

\begin{figure}
\begin{center}\includegraphics[width=1\columnwidth]{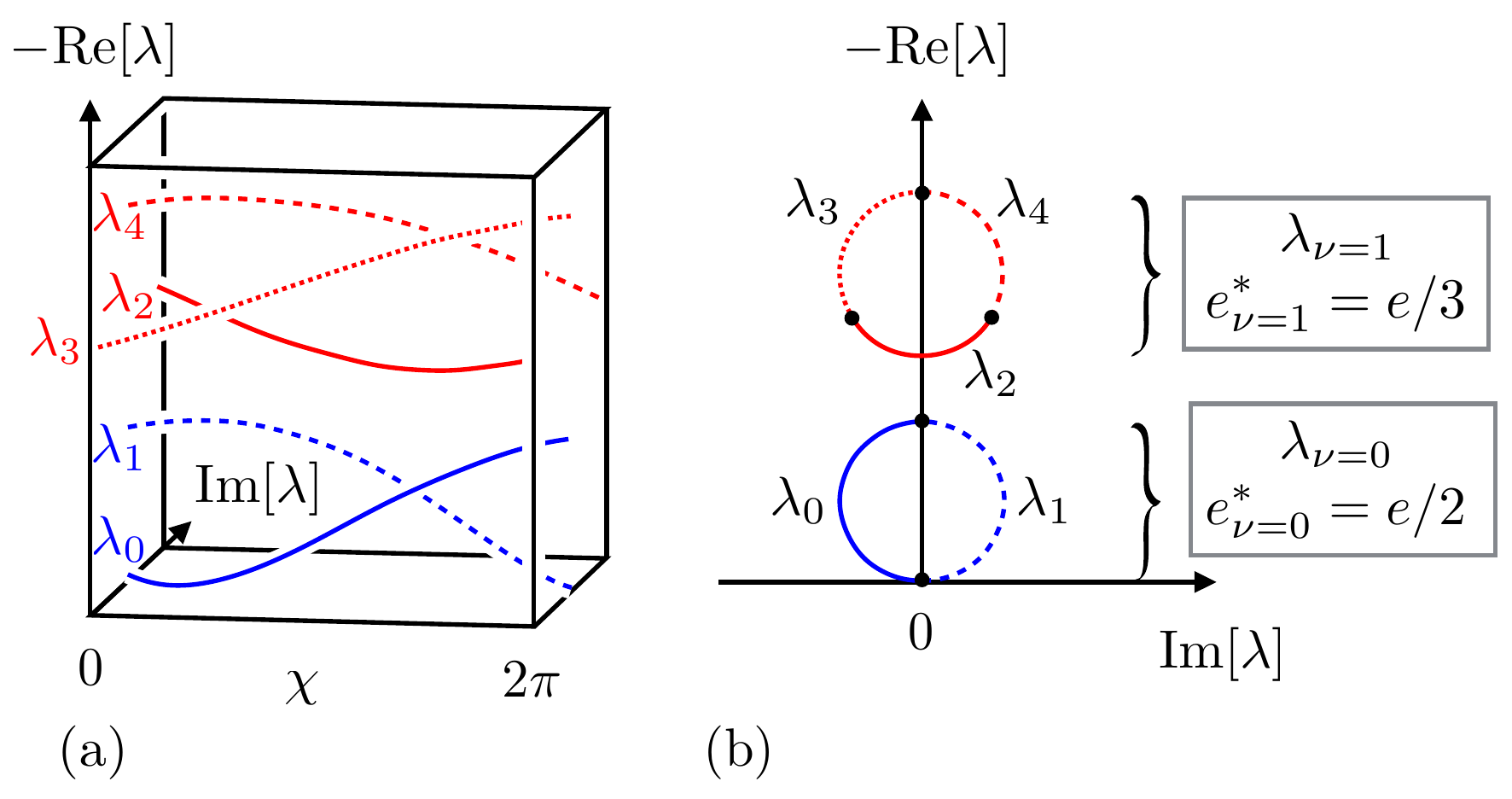}\end{center}

\caption{(a) Generic spectrum of $W\left(\chi\right)$, describing the combined dynamics of system and detector (see Sec.~\ref{sec:braiding_and_frac_charge}). There are two braid subblocks
with broken periodicity. The lower one (blue) has a periodicity of
$4\pi$, and the upper one (red) a periodicity of $6\pi$. (b) The
same spectrum projected onto the complex plane. The black dots indicate
the eigenvalues at $\chi=0$. We see that the five eigenvalues merge
into two bands, with the indices $\nu=0$ and $\nu=1$.
As explained in Sec.~\ref{sec:braiding_and_frac_charge}, we can assign
the charges $e/2$ and $e/3$ to these two bands. \label{fig_generic_spectrum}}
\end{figure}

Here, we present the novel idea that the notion of transport with fractional charges is by no means a unique property of systems with strong quantum correlations, and on the contrary can occur already in the simplest possible, classical transport regime of sequential electron tunneling. 
As we show, this fractionalization is a consequence of a topological
transition occurring in the full-counting statistics (FCS) of nonequilibrium transport, which can be described by the braid group \cite{Ren_2012,Li_2013}, referred to as a braid phase transition. 
Importantly, this transition occurs already for standard transport
through very simple systems, such as quantum dots or metallic islands. We show that each eigenmode of the topological detector dynamics contributes to the transport with an individual fractional charge, see Fig.~\ref{fig_generic_spectrum}.
We stress that while the fractionalisation mechanism we consider here is quite different, we find that its signatures in the low-frequency transport statistics are \textit{indistinguishable} from those occurring in the transport statistics of quasiparticles in strongly correlated systems~\cite{Kane_1994,Saminadayar_1997,de-Picciotto_1997,Reznikov_1999,Comforti_2002,Dolev_2008,Gutman2010}. Furthermore, by means of nontrivially quantized geometric phases emerging in the detector space, we show that the fractional charge effect in sequential electron tunneling represents a classical analogy to the fractional Josephson effect emerging in topological superconducting junctions~\cite{Fu_2009,Bocquillon_2016,Wiedenmann_2016,Deacon_2017,Zhang_2014,Orth_2015,Vinkler-Aviv_2017}.

\subsection{Fractional charges versus integer charge quantization}

In order to make our result appreciable, we first need to establish what is actually meant by a fractional charge. Obviously, we do not claim that transport via sequential electron tunneling literally splits the electron.
However, neither do strong quantum correlations. If we partition \textit{any} electronic system along a chosen sharp interface, we will measure an integer number of electrons on either side, irrespective of the interactions and the topology present in the system. Considering the FCS of transport, this fundamental limit is embedded in a \textit{global} property of the moment generating function $m(\chi)$ depending on the counting field $\chi$. Namely, $m$ is necessarily $2\pi$-periodic in $\chi$, as $\chi$ denotes the Fourier transform variable of the integer number of transported charges $N$.

The above statement goes however against the predictions of field theories where a non-integer valued charge operator emerges (see, e.g., \cite{Pham_2000}). Such theories provide fractional charges up to arbitrary statistical moments. This leads us to the question, how the notion of fractionally charged quasiparticles can be reconciled with the fundamental integer charge quantization? The answer lies in the \textit{local} properties of the moment generating function, respectively, in the analytic continuation of $m$. We can derive the individual cumulants of transport, such as the current $I$, the noise $S$, as well as the higher orders, from an expansion of $m(\chi)$ around $\chi=0$. In the simplest possible case of Poissonian transport, where the emission of individual excitations mediating
the current is uncorrelated, the charge $e^{*}$ of a quasiparticle can be extracted using only the lowest cumulants. Namely, one finds for the Fano factor
\begin{equation}
F\equiv\frac{S}{e\left|I\right|}=\frac{e^{*}}{e}\ .\label{eq:Fano_factor}
\end{equation}
This was
shown for the fractional quantum Hall effect \cite{Kane_1994,Saminadayar_1997,de-Picciotto_1997,Reznikov_1999,Comforti_2002,Dolev_2008},
and for the uncorrelated emission of Cooper pairs \cite{Jehl_2000}.
Very recently, the occurence of a universal fractional Fano factor
has been predicted also for transport through a charge Kondo device
\cite{Landau_2017}. 

Beyond the Poissonian limit, and hence beyond the lowest two cumulants, the notion of quasiparticle charges can be generalized as follows. We take Ref.~\cite{Gutman2010} as inspiration, where the FCS for transport through Luttinger liquid nanowires was derived. In that work, the authors show that within the framework of Luttinger liquid theory, the $2\pi$-periodicity of the moment generating function $m$ is indeed broken. As the authors explain, the origin of this violation is due to discarding modulations of the charge density on the length scale of the Fermi wave length - a standard procedure to arrive at Luttinger liquid theory~\cite{Haldane_1981}.
If therefore the transport statistics are measured across a sharp interface (sharp with respect to the Fermi wavelength) this literal fractionalization of charge cannot be physical. The authors assert that in this case, the moment generating function with broken periodicity is only correct for $\chi$ in between $-\pi$ and $\pi$, and beyond this interval, $m$
has to be continued periodically - thus reinstating charge quantization at the expense of discontinuities in $m(\chi)$. The fractional charge of the Luttinger quasiparticles remains nonetheless well-defined, as the analytic continuation of the moment generating function beyond the $-\pi$ to $\pi$ interval. To summarize, the result of~\cite{Gutman2010} suggests that the picture of fractionally charged quasiparticles is perfectly valid locally in $\chi$, that is, when expanding the moment generating functions into cumulants, whereas care has to be taken with respect to the \textit{global} properties of the moment generating function. In fact, this nontrivial relation between local and global properties of observables along a certain base space (here the counting field $\chi$) already indicates that topological considerations may become important.

In the here considered topological FCS of sequential electron tunneling, in particular for long measurement times, we will find the exact same feature of a fractional charge extractable from the analytic continuation of the moment generating function $m$, and a globally conserved $2\pi$-periodicity including discontinuities at $\chi=\pm \pi$. We note that in the context of Master equations, a nonanalytic behaviour of the moment generating function due to dynamical phase transitions has already been considered~\cite{Ren_2012,Flindt_2013,Hickey_2014}. Here we provide the crucial new insight, that these discontinuities indicate the presence of fractional charges, being at odds with integer charge quantization.

We then strongly generalize this finding to finite measurement times. The system-detector time evolution is described through a kernel $W(\chi)$ which has a set of eigenmodes as a function of $\chi$, as depicted in Fig.~\ref{fig_generic_spectrum}. There
is one stationary mode ($\nu=0$) corresponding to the unique stationary
state of the open system, and in general many decaying modes ($\nu>0$)
describing the relaxation dynamics. While $W$ remains $2\pi$-periodic in $\chi$ due to charge quantization, the underlying eigenspectrum may have a broken periodicity. We will show that in a generic case of a topological spectrum, we can map the FCS of ordinary electrons to a system transporting fractionally charged quasiparticles, observed with a detector resolving only integer charges. Here, we thus interpret the elementary charge $e$ as a fundamental \textit{resolution limit} of the transport or charge detector.  Each eigenmode $\nu$ contributes to transport with a different fractional charge, $e_{\nu}^{*}=e/p_{\nu}$, where the integer $p_{\nu}>1$ denotes
the broken periodicity, as shown
in Fig.~\ref{fig_generic_spectrum}. With this treatment we provide a simple generic mechanism for the topological transition: a detector with a finite resolution observing a discrete process with a sub-resolution step size. Of course, the quasiparticles appearing in our analogy are hypothetical, in the sense that the system is only physical once charge quantization is reestablished through the detector resolution. However, due to integer charge quantization being fundamental, we do not believe that they are more hypothetical than fractionally charged quasiparticles in other theories, such as Luttinger liquids~\cite{Pham_2000,Gutman2010} or similarly chiral Luttinger liquids describing the edge state of the FQHE~\cite{Wen1990}. Again, our argument relies on the fact that field theories containing quasiparticles with a non-integer  charge eigenvalue seem to fail to correctly reproduce the global properties of $m$~\cite{Gutman2010}. Unfortunately, a complete comparison of the two kinds of quasiparticles is currently out of reach: the argument to reinstate integer charge quantization in~\cite{Gutman2010}, in spite of being very plausible, appears without formal derivation. While generalizations of Luttinger liquid theory have been worked out~\cite{Imambekov2012}, we are not aware of an extension that preserves integer charge quantization~\cite{Schmidt_PC_2019}. A closer comparison of the here considered effect to strongly correlated systems is however possible for the special case of the fractional Josephson effect, as we explain in more detail below.
 
Concerning the ongoing quest in the community to detect exotic quasiparticles via their fractional transport properties~\cite{Keeling_2006,Jonckheere_2005,Glattli_2017,Kapfer2018Jun}, we can interpret our result with either a positive or a pessimistic ring. On the upside, we show that one can realize fractional charges in condensed matter systems with very simple means of ordinary classical transport, avoiding difficulties related to controlling strongly correlated quantum systems. It is merely the mechanism for creating the fractional charges which is different, as in our case, all we need is a topological transition in a completely classical dynamics, where the maximal fractionalization depends simply on the number of available, local degrees of freedom.

However, regarding the effort to experimentally verify the existence of exotic fractional excitations such as Laughlin quasiparticles, our work suggests that signatures in the transport statistics, at the very least in the here considered low-frequency regime, may not be as unique as generally expected. We note that the context of the integer quantum Hall effect, it has already been predicted, that so-called ``half-Levitons'' can be generated even in the absence of strong correlations~\cite{Moskalets2016}. In that work it is however required that the flux induced by the injection voltage equals exactly $\pi$, leaving thus the open question, how far the effect is protected. The mechanism we propose on the other hand is of topological nature, and therefore does not require a fine-tuning of parameters. 

\subsection{Geometric phases and classical analogy to fractional Josephson effect}

As we argue here, integer charge quantization represents a crucial ingredient in understanding the transport statistics with fractional charges. In the second part of this paper, we show that the incompatibility of fractionally charged quasiparticles and integer charge quantization results in a nontrivially quantized geometric phase defined along the counting field. In fact, we find that the geometric phase provides an additonal topological number in its own right. It does not merely reflect the topology of the eigenspectrum, but relates the periodicity of the eigenspectrum (fractional charges) to the periodicity of the operator generating the transport dynamics (integer charge quantization).

Since the system-detector dynamics are described by a non-hermitian kernel, this result connects to recent efforts in defining geometric phases and topological numbers in the context of non-hermitian Hamiltonians describing optic systems~\cite{Leykam_2017,Kunst_2018,Edvardsson_2019}, and contributes to the fundamental open question of generalizing
geometric phases and related topological numbers to open, nonequilibrium
quantum systems \cite{Bardyn_2013,Budich_2015,Pluecker_2016,Pluecker_2017,Leykam_2017,Kunst_2018,Bardyn_2018,Edvardsson_2019} by explicitly proposing topological numbers defined in the detector space.

Importantly, by means of this geometric phase, we can show that the charge fractionalization in sequential electron transport can be considered as a classical analogy to the fractional Josephson effect, emerging in topological Josephson junctions~\cite{Fu_2009,Bocquillon_2016,Wiedenmann_2016,Deacon_2017,Zhang_2014,Orth_2015,Vinkler-Aviv_2017}. Here, the classical counting field $\chi$ has its counterpart in the phase difference $\phi$ describing the quantum coherent superconducting transport across Josephson junctions. Just as a $2\pi$-periodicity in $\chi$ signifies that the electron charge $e$ is the underlying unit of incoherent transport, a $2\pi$-periodicity in $\phi$ indicates the Cooper pair charge $2e$ as the fundamental unit of superconducting transport. In topological Josephson junctions, the usual $2\pi$-periodicity
in the superconducting phase difference $\phi$ gets broken due to
the transport of ``fractional'' Cooper pairs, such as the tunneling
of Majorana- or parafermions with a $4\pi$ \cite{Fu_2009,Bocquillon_2016,Wiedenmann_2016,Deacon_2017},
respectively an $8\pi$ Josephson effect \cite{Zhang_2014,Orth_2015,Vinkler-Aviv_2017}.
However, we have to insist again, that when measuring supercurrents across a sharp interface, it is only the eigenspectrum describing the supercurrent transport which exhibits a broken periodicity. The Hamiltonian describing the supercurrent transport should remain $2\pi$-periodic, indicating that the Cooper pair still is the fundamental unit in which the supercurrent is transported and eventually measured. In fact, as we already pointed out, we believe that in Luttinger liquids a formal requantization of charge is highly nontrivial; so far we have only encountered qualitative arguments, as in Ref.~\cite{Gutman2010}. For topological superconducting junctions on the other hand, we will argue that fixing the global $2\pi$-periodicity of the Hamiltonian as a function of $\phi$ can be formally justified in a straightforward fashion. Thus, we can show that our result of quantized geometric phases can be easily transferred to supercurrents in topological Josephson junctions by replacing $\chi$ with $\phi$. In this way, we are able to establish a deep analogy between fractional charges in strongly correlated systems and in classical sequential electron tunneling.

In fact, we can think
of the fractional transport of electrons as a classical simulation
of the fractional Josephson effect. In a similar spirit, the simulation of topological
features known from quantum coherent systems by means of a classical
stochastic dynamics has been recently and prominently demonstrated
in the diffusion of polymers \cite{Souslov_2017,Abbaszadeh_2017},
by exploiting the structural similarity between the Schrödinger equation
and the diffusion equation. Likewise, the realization of a Su-Schrieffer-Heeger
(SSH) model has been recently proposed using the full-counting statistics
of single electron transistors \cite{Engelhardt2017}. We note that we have reason to believe that in our particular case, the simulation might actually be more stable than the original. The fractional Josephson effect is strongly susceptible
to the breaking of fermion parity \cite{Fu_2009,vanHeck_2011,Rainis_2012}.
The topology in the incoherent transport on the other hand is \textit{defined}
in the single electron transport, and thus by nature stable with respect
to parity breaking.

\subsection{Experimental verification and possible applications}

In the final part of our work, we aim to point towards experimental verifications of our claims. First, we study a nearly Poissonian regime, where a measurement
of the lowest few cumulants of the current suffices to determine the
fractional charge of the stationary mode, along similar lines as Eq.~(\ref{eq:Fano_factor}).
By means of the topological argument, we show that the shot noise
regime in quantum dots, which was hitherto considered as strictly
sub-Poissonian~\cite{Thielmann_2003}, should actually be interpreted as a nearly Poissonian transport with a fractional charge. With the continued
experimental progress in the detection of higher cumulants of the
current \cite{Reulet_2003,Bomze_2005,Gustavsson_2006,Fujisawa_2006,Timofeev_2007,Sukhorukov_2007,Gershon_2008,LeMasne_2009,Flindt_2009,Ubbelohde_2012,Brandner_2016}
we believe that the fractional nature of sequential electron transport could be verified
with existing technology. 

When going beyond this Poissonian regime, and when trying to measure fractional charges of decaying modes, we need to
have access to the time-dependent dynamics of the detector. These can be measured by means of
the detectors waiting time distribution, describing the statistics
of time intervals between transport events \cite{Koch_2005,Brandes_2008,Welack_2009,Albert_2011,Rajabi_2013,Sothmann2014,Potanina_2017}.
Importantly, we are able to show that the waiting times can also provide the geometric phases.
We study easily realizable models, where the waiting times are accessible
through time resolved measurement of the charge state of the system, and provide some recipes for acquiring the geometric phases. At the end, we study some realistic
cases of detector errors and comment on detector backaction effects. We argue that neither of them can fundamentally hamper the measurement of the topological features.

Finally, through a simple extension of the studied models, we briefly show that braid phase transitions in the detector dynamics may open up unexpected applications for new measurement techniques going beyond the detector resolution. Namely, such transitions occurs likewise, when considering a detector which is not necessarily restricted by some fundamental limit, but rather due to a deficiency.

This paper is structured as follows. In Sec.~\ref{sec:braiding_and_frac_charge}
we establish that the braid topology corresponds to fractional charges in classical, stochastic transport and perform a comparison to FCS in Luttinger liquids. We then show in Sec.~\ref{sec:fractional_josephson_effect}
the emergence of quantized geometric phases defined along the counting field, which depend on the ratio between the fractional charges and the integer charge quantization. In the same section, we demonstrate the analogy to the fractional Josephson
effect, known from topological superconducting junctions. Finally, in Sec.~\ref{sec:Strategies-for-experiments} we provide strategies to verify our claims in experiment.

\section{Topology in full counting statistics and fractional charges\label{sec:braiding_and_frac_charge}}

\subsection{Transport statistics and braid group\label{subsec:braiding_in_FCS}}

In this work, we focus on sequential tunneling of electrons through
small quantum systems, such as quantum dots or metallic islands. Let us first provide the main assumptions concerning the dynamics
of the open quantum system in the absence of transport measurement.
Namely, we assume a local system weakly tunnel coupled to two (or more) reservoirs. With weak coupling, we here mean that the time scale at which the electrons tunnel is slower than the time scale at which correlations in reservoirs decay. Given the tunneling rate $\Gamma$ (for a definition see Sec.~\ref{subsec:explicit_examples} and Sec.~\appsinglelevelQD), the weak coupling picture is valid as long as $\Gamma\ll \tau_\text{res}^{-1}$, where the inverse of the time scale at which correlations decay in the reservoirs. The latter scales with the dominant energy scale given by the reservoir. Close to resonance $\tau_\text{res}^{-1}$ scales with the reservoir temperature. When the  chemical potentials of the reservoirs are strongly detuned from the systems energy states, then $\tau_\text{res}^{-1}$ scales with the difference between said energies and the chemical potential.

Through tracing out the reservoir degrees of freedom, the state of
the local system is described by a reduced density matrix. We focus
on the simplest possible case, when the dynamics of the system is
completely incoherent, such that the reduced density matrix stays
diagonal in a given basis (in our case, the charge and energy eigenbasis of the quantum system). In this case, we may discard the offdiagonal
elements, and capture the diagonal part of the density matrix as a
vector of probabilities, $\left|P\right)$. In the weak coupling regime, the time evolution of $\left|P\right)$ is given by a fully
markovian master equation
\begin{equation}
\partial_{t}\left|P\left(t\right)\right)=W\left|P\left(t\right)\right)\ ,\label{eq:dynamics_no_detector}
\end{equation}
where $W$ can be represented as a trace preserving matrix, with a set of eigenvalues
$\left\{ \lambda_{n}\right\} $. It can be decomposed into a sum of contributions, $W=\sum_\alpha W_\alpha$, where each $W_\alpha$ describes the coupling to one reservoir $\alpha$. We focus on matrices $W$ with a
unique stationary state, with eigenvalue $\lambda_{0}=0$, while the
other (generally nondegenerate) eigenvalues $\lambda_{n\neq0}$ have
a real part $<0$. We refer to the corresponding eigenmodes with
$n\neq 0$ as decaying modes. We furthermore focus on matrices $W$, which
have in addition corresponding left and right eigenvectors, $\left|n\right)$
and $\left(n\right|$, where the notation is such that any $\left|a\right)$
is a vector, whereas any $\left(b\right|$ should be regarded as a
map $\left(b\right|\cdot$ from a vector to a scalar. The right eigenvector
to $\lambda_{0}=0$, $\left|0\right)$, corresponds to the density
matrix of the stationary state, $\left|0\right)=\lim_{t\rightarrow\infty}\left|P\left(t\right)\right)$.
The corresponding left eigenvector $\left(0\right|$ is simply the
operator tracing over the degrees of freedom of the quantum system,
$\left(0\right|\cdot=\text{tr}_{S}\left[\cdot\right]$. Thus, $(0|$ having eigenvalue $0$ expresses
the trace preserving property of $W$, i.e., $(0|W=0$. The decaying modes $n\neq0$
may likewise have a specific physical interpretation, depending on
the system under consideration. For quantum dot systems (as we consider
them later) such higher modes correspond to the decay of the charge
or spin of the quantum system \cite{Splettstoesser_2010}, or even
a quantity related to fermion parity \cite{Contreras-Pulido_2012,Saptsov_2012,Saptsov_2014,Schulenborg_2014}.

Now we come to the measurement of transport. For for the sake of simplicity, but without loss of generality, we limit our considerations to two reservoirs, one left and one right, such that we may write $W=W_\text{L}+W_\text{R}$. The central object of interest is the probability $p(N,\tau)$ of having transported $N$ electrons into a given reservoir, after a measurement time $\tau$. The moment generating function is then defined as the Fourier transform of this probability,
\begin{equation}\label{eq_m_of_N}
m(\chi,\tau)\equiv \sum_{N}e^{i\chi N}p\left(N,\tau\right)\ ,
\end{equation}
where $\chi$ is referred to as the counting field.

We describe the FCS by the addition of an explicit detector with a degree of freedom $N$, coupled to the system such that $N\rightarrow N\pm1$ when an electron exchange is measured. The detector is thus part of the total physical system storing the information of the transport processes~\cite{Levitov_1996}, and the dynamics of system plus detector can be described in terms of an all-encompassing Hamiltonian operator~\cite{Schaller_2009,Pluecker_2016,Pluecker_2017}. Evaluating $m(\tau)$ consequently corresponds to a projective measurement of the detector state at time $\tau$. Taking into account the composite detector and system degrees of freedom, we receive $\left|P\left(t\right)\right)\rightarrow\left|P\left(N,t\right)\right)$
and $W\rightarrow W\left(N-N'\right)$. The off-diagonal elements in $W$ correspond to all processes that transport a nonzero charge
$N-N'\neq 0$. The resulting dynamics of system \textit{and} detector
is now given as $\partial_{t}\left|P\left(N,t\right)\right)=\sum_{N'}W\left(N-N'\right)\left|P\left(N,t\right)\right)$. We perform again the same Fourier transform, resulting in the equation,
\begin{equation}
\partial_{t}\left|P\left(\chi,t\right)\right)=W\left(\chi\right)\left|P\left(\chi,t\right)\right)\ .\label{eq:dynamics_system_and_detector}
\end{equation}
By means of this Fourier transform, it is easy to understand the relation between $W\left(\chi\right)$ (for the system-detector
dynamics) and $W=W\left(0\right)$ (for the system dynamics only). Namely, $\chi=0$ corresponds simply to tracing over
the detector degrees of freedom $N$, thus discarding the information stored in the detector
state, and recovering the system dynamics without the detector.

Equation~\eqref{eq:dynamics_system_and_detector} establishes the combined system-detector dynamics. Note that these dynamics are valid when considering time scales $\tau$ slower than the reservoir correlation time, $\tau_\text{res}$. We refer to this as the low-frequency limit of FCS. Note however that due to weak coupling, $\Gamma\ll \tau_\text{res}^{-1}$, we can not only correctly describe the zero-frequency limit of FCS $\tau^{-1}\ll\Gamma$ (which we also refer to as the long measurement time limit), but also a finite frequency regime, where $\tau^{-1}$ is not limited with respect to $\Gamma$ (as long as $\Gamma,\tau^{-1}\ll\tau_\text{res}^{-1}$). Based on these dynamics, we are able compute the moment generating function after a measurement time $\tau$ [see Eq.~\eqref{eq_m_of_N}] as,
\begin{align}
m\left(\chi,\tau\right) & =\left(0\left(0\right)\right|e^{W\left(\chi\right)\tau}\left|0\left(0\right)\right)\label{eq:cumulant}\\
 & =\sum_{n}\alpha_{n}\left(\chi\right)e^{\lambda_{n}\left(\chi\right)\tau}\ ,\label{eq:cumulant_eigenvalues}
\end{align}
where we expressed the right-hand side explicitly
in terms of the eigenspectrum of $W(\chi)$. That is, $\alpha_{n}\left(\chi\right)=\left(0\left(0\right)\left|n\left(\chi\right)\right)\right.\left(n\left(\chi\right)\left|0\left(0\right)\right)\right.$, where $\lambda_n(\chi)$ and $|n(\chi)$ as well as $(n(\chi)|$ are the eigenvalues, and right as well as left eigenvectors of $W(\chi)$. The labels of all $n$ at finite $\chi$ are chosen such that they correspond the same labels for $\chi=0$. Consequently, with the eigenvalue $\lambda_{0}\left(\chi\right)$
we denote the mode that corresponds to the stationary state for $\chi=0$,
such that $\lambda_{0}\left(0\right)=0$. In Eq.~\eqref{eq:cumulant_eigenvalues} we assumed the system to be in the stationary state $|0(0))$ initially, which is however not a necessary condition for our following results. We note furthermore that we here consider
systems where $W$ can be always decomposed into left and right eigenvectors for real $\chi$. In Sec.~\appone, we do however show that there
are special, isolated points (commonly referred to as exceptional points) in the space of complex $\chi$, where
a decomposition into eigenvectors is not possible. In fact, these points generate the
here considered topology (see later in this section, as well as Sec.~\appone). 

Up until now, we have considered the detector as an actual physical quantity, with a measurable state. There is an alternative way of formulating and interpreting the FCS, in terms of the current operator $\widehat{I}$ as the observable, i.e., through the expectation value of $\widehat{I}$ and higher cumulants (current noise, skewness, and so forth). As we will see, there are some subtle differences between the two notions of FCS in the topological phase. They can be related as follows. One defines the cumulant generating function $c$ as $m(\chi)=e^{c(\chi,\tau)\tau}$. Then, the cumulants are obtained through a local Taylor expansion of $c$ around $\chi=0$,
\begin{equation}
C_{k}(\tau)=\left.\left(-ie\right)^{k}\partial_{\chi}^{k}c\left(\chi,\tau\right)\right|_{\chi\rightarrow0}\ .\label{eq:cumulants_def_2}
\end{equation}
For instance, the current expectation value is given as $I=C_{1}$, or the current noise as $S=C_{2}$.  Thus, the Taylor series of the cumulants provides the analytic continuation of the moment generating function. 
Finally, in the limit of very long measurement times $\tau\rightarrow\infty$, one can easily see in Eq.~\eqref{eq:cumulant}, that only the eigenvalue with the lowest real part survives. Thus, for the cumulant generating function, at least locally at $\chi$ close to zero, one finds~\cite{Bagrets_2002,Bagrets_2003}
\begin{equation}\label{eq_cumulant_long_times}
c_{\infty}\left(\chi\right)\equiv\lim_{\tau\rightarrow\infty}c\left(\chi,\tau\right)=\lambda_{0}\left(\chi\right)\ .
\end{equation}
This limit can be considered as a thermodynamic limit of a macroscopic number of transported electrons.

Importantly, returning now to the picture of an explicit detector, we have so far only specified that we measure the transport into a given reservoir. But we have not yet specified which one of the reservoirs. Moreover, due to current conservation, the stationary current entering from the left reservoir must be the same as the entering the right reservoir. Hence, we could also consider an appropriately weighted sum of transport measurement at both reservoirs, as long as their sum still provides the same current. Such a specification seems like a triviality, especially in the long measurement time limit, where the eigenvalue $\lambda_0(\chi)$ alone enters in the observables, see Eq.~\eqref{eq_cumulant_long_times}. Namely, as we argue in a moment, there are many different detector setups providing the same eigenvalues, precisely thanks to current conservation. However, in the remainder of this work we will encounter some subtleties which render this question nontrivial. Namely, unlike the eigenvalues, the \textit{eigenvectors} are sensitive to the measurement setup, and they enter the observables for finite times $\tau$, or when discussing geometric phases (see Sec.~\ref{sec:fractional_josephson_effect}).

To appreciate the importance of the measurement setup, let us start by showing that to ensure integer charge quantization, it will be necessary to choose a sharp interface, across which transport is measured. For the here considered small quantum systems (such as quantum dots), this leaves us with two possibilities. Either we place the detector at the interface between the quantum system and the right, or the left contact. For a detector attached to the right (which will be the default setting for the rest of the paper, unless specified differently), the kernel including counting fields can be represented in a compact way as
\begin{equation}\label{eq_W_chi_right_detector}
W(\chi)=W_\text{L}+e^{-i\chi\widehat{n}}W_\text{R}e^{i\chi\widehat{n}}\ ,
\end{equation}
where $\widehat{n}$ is a superoperator, defined such that $(0|\widehat{n}|P)$ returns the charge expectation value of the quantum system. We will define $\widehat{n}$ explicitly later, when dealing with explicit models. If we were to measure at the interface to the left reservoir on the other hand, we would find a different kernel, which can be obtained from the above through the unitary transformation, $e^{i\chi\widehat{n}}W(\chi)e^{-i\chi\widehat{n}}$. Both of these kernels are manifestly $2\pi$-periodic, since $\widehat{n}$ has integer eigenvalues, thus ensuring the $2\pi$-periodicity of $m(\chi)$ [see Eq.~\eqref{eq:cumulant}], and preserving integer charge quantization. To emphasize the relation between charge quantization and sharp interfaces, let us now consider an example, where the detector is, in some sense, not precise. Imagine a detector that measures currents out of the left and into the right contact, and superposes them with a probability of $p$ and $1-p$, respectively. Such setups are likewise described through a unitary transformation, $e^{i\chi\widehat{n}p}W(\chi)e^{-i\chi\widehat{n}p}$. Because they are related to a unitary transformation, all these kernels will provide the same eigenvalues as the kernel in Eq.~\eqref{eq_W_chi_right_detector}, and consequently the same cumulants in the long time limit (due to the aforementioned current conservation). However, for the unprecise detection schemes, $p\neq 0,1$, the kernels have a broken $2\pi$ periodicity, and thus violate integer charge quantization. This is however not unphysical (similar to the argument in~\cite{Gutman2010}), because we do not measure across a single sharp interface, but at two interfaces, with a statistical uncertainty $\sim p(1-p)$. This indicates at a very simple example the necessity to measure at a sharp interface.

Now, we discuss the topology of the eigenspectrum of $W(\chi)$, as first studied by Ren and Sinitsyn~\cite{Ren_2012}. This notion of topology arises from two ingredients. The first ingredient is the $2\pi$-periodicity
of $W\left(\chi\right)$ in $\chi$ due to the discreteness of the transport process (when measured across a sharp interface), from which follows that
at a certain value of $\chi$, $W(\chi)$ has the same set of eigenvalues as at $\chi+2\pi$. The
second ingredient is the continuity of the eigenspectrum as a function of $\chi$,
in the sense that locally (in $\chi$) we can always find a labelling $n$
of eigenvalues, such that $\partial_{\chi}\lambda_{n}\left(\chi\right)$
is well-defined. This allows us to examine, how the eigenvalues are
connected in a $2\pi$ interval in $\chi$. And as shown in \cite{Ren_2012} this connection can be nontrivial, such that the eigenspectrum
of the kernel $W\left(\chi\right)$ can undergo topological transitions,
which are characterized according to the braid group \cite{Artin_1947}.

In fact, as we foreshadowed above, we are able to relate the topology of the eigenspectrum along $\chi\in\mathbb{R}$ to exceptional points in the space of a complex counting field $\chi\in\mathbb{C}$, i.e., isolated points, where certain eigenvalues are degenerate and the corresponding eigenvectors are ill-defined (for more details, see Sec.~\appone). To each of the exceptional points, we can assign a generator of the braid group. Thus, depending on the configuration of the exceptional points in the space of the complex counting field, the topology of the eigenspectrum for real $\chi$ is described by a certain element of the braid group. This provides us in some sense with a bulk-boundary correspondance. In the 2D space of complex counting fields there emerge special points (exceptional points) which carry a braid generator (similar to a topological charge), which then in turn define the topology of the eigenspectrum along the 1D space of the real counting field. Transitions between different topologies occur through a passing of an exceptional point across the real axis of $\chi$. This provides us furthermore with a concrete argument for the topological protection of the effect. Namely, small variations in the kernel $W(\chi)$ give rise to only small shifts of the exceptional points. Thus, the topology of the eigenspectrum is stable within a connected parameter subspace. Moreover, as we show in Sec.~\appone, for generic kernels $W$ there is no fundamental symmetry (apart from equilibrium transport, see also later) which forbids the presence of exceptional points, and thus a braid phase transition is always possible, in a generic nonequilibrium setup. The exceptional points will again be important when discussing geometric phases in Sec.~\ref{sec:fractional_josephson_effect}.

Of particular interest in this work is the subgroup of braids whereby
the $2\pi$-periodicity of the kernel $W\left(\chi\right)$ can be broken on the level of the underlying eigenspectrum.
Namely, different eigenvalues can wind around each other when continuously
advancing $\chi$ by $2\pi$, such that $\lambda_{n}\left(\chi+2\pi\right)=\lambda_{n'}\left(\chi\right)$
with $n'\neq n$. We define the periodicity $p_{n}\in\mathbb{N}$
for a particular eigenvalue $n$ to return to its original value through
$\lambda_{n}\left(\chi+2\pi p_{n}\right)=\lambda_{n}\left(\chi\right)$,
i.e., as the integer number of times we need to shift $\chi$ by $2\pi$
to return to the original eigenvalue. Importantly, from the $2\pi$-periodicity of $W(\chi)$, it follows that the eigenvectors
satisfy the same relations, such that $\left|n\left(\chi+2\pi p_{n}\right)\right)\left(n\left(\chi+2\pi p_{n}\right)\right|=\left|n\left(\chi\right)\right)\left(n\left(\chi\right)\right|$.
We provide a generic example of a system with 5 eigenvalues in Fig.
\ref{fig_generic_spectrum}a. In this example, we chose a spectrum
which is comprised of two braid subblocks that have a different
periodicity. In the example of Fig.~\ref{fig_generic_spectrum}, $\lambda_{0}\left(\chi+2\pi\right)=\lambda_{1}\left(\chi\right)$
and $\lambda_{1}\left(\chi+2\pi\right)=\lambda_{0}\left(\chi\right)$,
as well as $\lambda_{2}\left(\chi+2\pi\right)=\lambda_{3}\left(\chi\right)$,
$\lambda_{3}\left(\chi+2\pi\right)=\lambda_{4}\left(\chi\right)$,
and $\lambda_{4}\left(\chi+2\pi\right)=\lambda_{2}\left(\chi\right)$. 

Finally, let us stress that, due to the overall $2\pi$-periodicity
of $W\left(\chi\right)$, the breaking of the periodicity on the level
of the eigenvalues and eigenvectors goes hand in hand with a \textit{redundancy}.
That is, the individual eigenvalues and eigenvectors in the same braid subblock are
merely shifted images of each other, and thus contain all the same
information. This fact will be of importance in Sec.~\ref{subsec:periodicity_and_charge}.
Considering the eigenvalues as a function of $\chi$ as generalized,
complex bands, and $\chi$ as a detector momentum~\footnote{In analogy to bands in solid state physics, where here, $\chi$ resumes a similar role as the $k$-vector in the Hamiltonian description of a crystal, i.e., a detector momentum.},
we may regard the braid phase transition with a breaking of periodicity as a merging
of bands. We account for the resulting redundancy by assigning a new
band index $\nu$, which does not count individual eigenvalues, but
enumerates individual subblocks (i.e., the actual independent bands). This is illustrated in Fig.
\ref{fig_generic_spectrum}b, where the spectrum has been projected
onto the complex plane. Here, the eigenvalues $n=0,1$ form the new topological
band $\nu=0$, and the eigenvalues $n=2,3,4$ form the band $\nu=1$.
In essence, enumerating the eigenvalues with the index $n$ is meaningful
when looking at the spectrum at a specific value of $\chi$, whereas the index
$\nu$ is meaningful for indexing the eigenspectrum when putting it into relation with the global $2\pi$-periodicity of $W$.

In the following, we will argue that the periodicity breaking due
to the nontrivial topology results in a transport statistics in terms of
\textit{fractional} charges in the same sense as it occurs in the above mentioned strongly correlated systems.  Each of the bands $\nu$ provides a separate fractional charge $e_{\nu}^{*}=e/p_{\nu}$. That is, in the
example given in Fig.~\ref{fig_generic_spectrum}, the band $\nu=0$
has a fractional charge of $e/2$, and the band $\nu=1$ contributes
to the transport with a charge $e/3$. In the next sections, we will provide
a careful argumentation for the fractional effect, and show that it
is, as a matter of fact, an extremely common occurrence in many simple
transport situations.

\subsection{Fractional charges of the stationary state \label{subsec:explicit_examples}}
We first want to establish the emergence of a fractional
charge in quantum dot systems for the stationary mode $n=0$, i.e., when focussing on a transport measurement in the limit of long measurement times $\tau\rightarrow\infty$. We first discuss the simplest generic transport model, the single-level quantum dot, where a braid phase transition occurs for the stationary state for a sufficiently strong out-of-equilibrium bias, leading to a fractional charge $e/2$. We then extend the model to a serial double quantum dot, where in addition a fractional charge $e/3$ can occur in the stationary state. For the latter model, we also discover topological phases with a trivial stationary state, but nontrivial decaying modes, requiring an understanding of fractional charges beyond the long measurement time limit.

\subsubsection{Single-level quantum dot}
Let us begin with the simplest possible model of a
single-level quantum dot coupled to two reservoirs. As we will argue now, there occurs a topological phase where
the transport is correctly described in terms of fractional charges. We then put this
result in relation with the FCS obtained in Luttinger liquid nanowires~\cite{Gutman2010}. Crucially, even though the two systems are physically very different, we will find that the moment generating function provides the \textit{same} signatures of fractional charges for both systems.

\begin{figure}
\begin{center}\includegraphics[width=0.9\columnwidth]{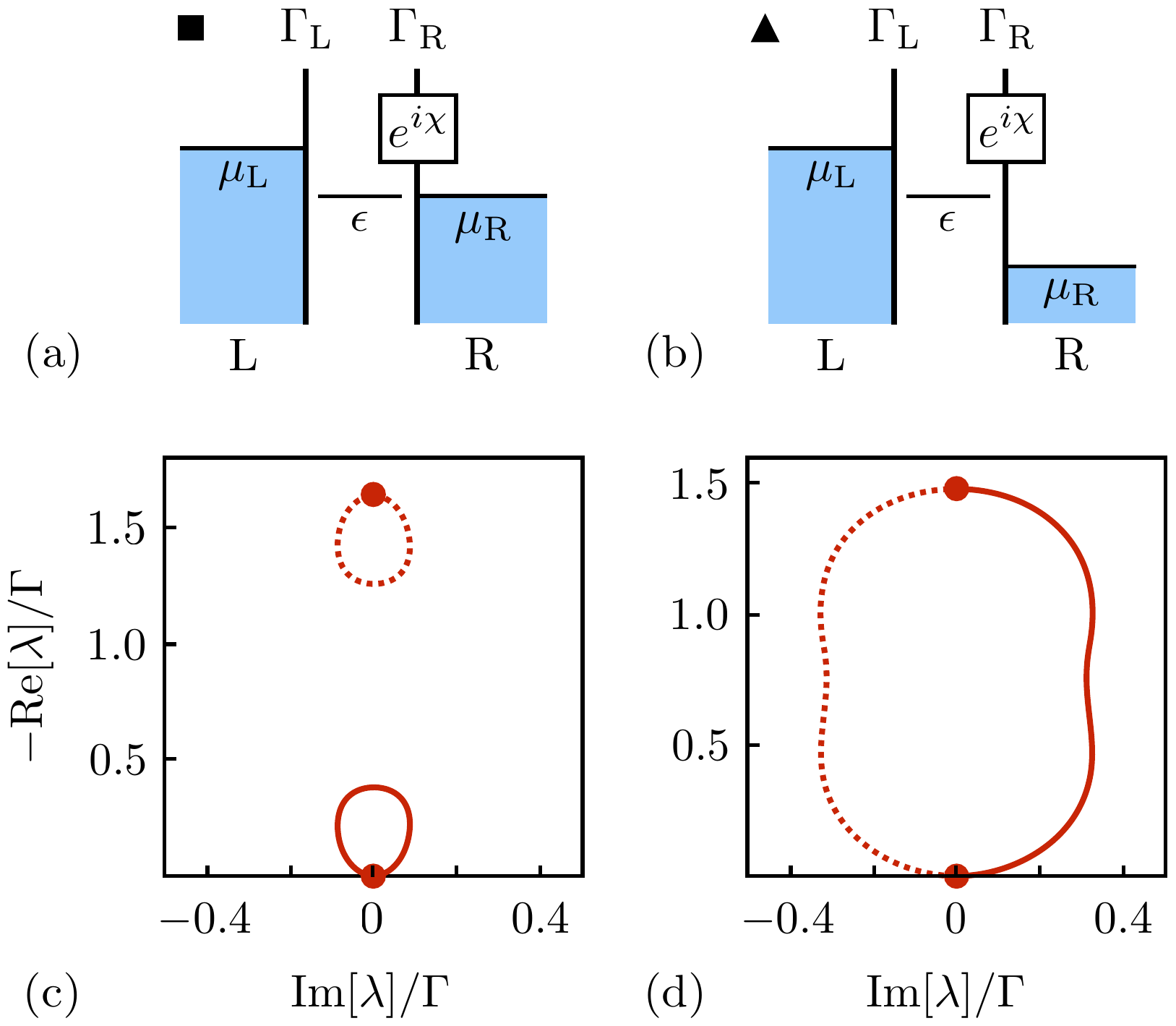}\end{center}

\caption{(a) Single-level quantum dot system, where the transport
statistics are measured between dot and right reservoir. The bias
and energy configuration correspond to the point marked with a square
in Fig.~\ref{figure_SQD_fractional_charge}a, where the transport
is trivial. (b) The same quantum dot system, however in a different
configuration, marked by a triangle in Fig.~\ref{figure_SQD_fractional_charge}a,
where the transport is in the topological phase. (c) and (d) Projection
of the complex eigenspectrum of $W\left(\chi\right)$, see Eq.~(\ref{eq:W2by2_eigenvalues}),
for the same parameters as in Fig.~\ref{figure_SQD_fractional_charge}a,
at the square point $\mu_{\text{L}}-\epsilon=k_{\text{B}}T$ and $\mu_{\text{R}}-\epsilon=0$
(c) and at the triangle point $\mu_{\text{L}}-\epsilon=k_{\text{B}}T$
and $\mu_{\text{R}}-\epsilon=-2k_{\text{B}}T$ (d). In (d) the spectrum
is braided in such a way, that the two eigenvalues have a $4\pi$-periodicy,
corresponding to a charge $e/2$. The solid points indicate
the eigenvalues for $\chi=0$.}
\label{figure_SQD}
\end{figure}

We assume that a single-level quantum dot has very low capacitances, such that the energy associated with charging the system is far
greater than the temperature. Let us set the energy of the single quantum dot level such that it is in the resonant configuration, where only a transition between
the empty and the singly filled level, $\left|0\right\rangle $ and
$\left|1\right\rangle $, is possible (we take into account the possibility
that the state $\left|1\right\rangle $ may be spin-degenerate). We
weakly couple the dot to two ordinary metallic reservoirs, one left and one
right, and we measure the transport statistics at the tunnel contact
to the right reservoir, see Figs.~\ref{figure_SQD}a and b. The full
Hamiltonian is described in Sec.~\appsinglelevelQD. 

The state of the quantum dot can be described by a (diagonal) density matrix, which reads, in vector form,
$\left|P\right)=\left(P_{0},P_{1}\right)^{T}$, where $P_{0}$ and
$P_{1}$ are the probability to be in either of the two charge states.
The corresponding kernel for weak tunnel coupling reads 
\begin{equation}
W\left(\chi\right)=W_{\text{L}}+e^{i\chi\widehat{n}}W_{\text{R}}e^{-i\chi\widehat{n}}\ ,\label{eq:kernel_single}
\end{equation}
with $\widehat{n}=\text{diag}\left[\left(0,1\right)\right]$ and \cite{Konig_thesis}
\begin{equation}
W_{\alpha}=\left(\begin{array}{cc}
-\sigma\Gamma_{\alpha}f_{\alpha} & \Gamma_{\alpha}\left[1-f_{\alpha}\right]\\
\sigma\Gamma_{\alpha}f_{\alpha} & -\Gamma_{\alpha}\left[1-f_{\alpha}\right]
\end{array}\right)\ ,\label{eq:W_alpha}
\end{equation}
where $\alpha=\text{L,R}$, and $\sigma$ indicates the degeneracy
of the quantum dot level, e.g., for the ordinary spin-degenerate case, $\sigma=2$.
The transition rates are given by the tunnling rates $\Gamma_{\alpha}$
and the Fermi functions, $f_{\alpha}=1/\left(1+e^{\beta\left[\epsilon-\mu_{\alpha}\right]}\right)$,
where $\beta$ is the inverse of the thermal energy $k_{\text{B}}T$,
and the energy differences $\epsilon-\mu_{\alpha}$ are controlled
by gate and bias voltages. We furthermore define the sum of tunneling
rates $\Gamma=\Gamma_{\text{L}}+\Gamma_{\text{R}}$. Importantly,
with the same kernel, we may also describe metallic islands instead
of quantum dots on a similar footing~\cite{Konig_thesis}. Therefore, while for
the remainder of this section we focus on the quantum dot case, we
stress that the results discussed below hold qualitatively also for
metallic islands.

The spectrum of $W\left(\chi\right)$ in Eq.~(\ref{eq:kernel_single})
has two eigenvalues, $\lambda_{0}\left(\chi\right)$ and $\lambda_{1}\left(\chi\right)$,
which are given as
\begin{equation}
\lambda_{0,1}\left(\chi\right)=-\frac{\gamma_\text{c}}{2}\left[1\mp\sqrt{1+u(\chi)}\right]\ ,\label{eq:W2by2_eigenvalues}
\end{equation}
with
\begin{equation}\label{eq_z_chi}
u(\chi)=r_+\left(e^{i\chi}-1\right)+r_-\left(e^{-i\chi}-1\right)\ ,
\end{equation}
and $\gamma_\text{c}=\sum_{\alpha}\left(\sigma\Gamma_{\alpha}f_{\alpha}+\Gamma_{\alpha}f_{\alpha}^{-}\right)$ is the charge relaxation rate, such that $\lambda_1(\chi=0)=-\gamma_\text{c}$ (see also Ref.~\cite{Splettstoesser_2010}), and $r_+=4\sigma\Gamma_\text{L}\Gamma_\text{R}f_\text{L}f_\text{R}^-/\gamma_\text{c}^2$ as well as $r_-=4\sigma\Gamma_\text{L}\Gamma_\text{R}f_\text{L}^-f_\text{R}/\gamma_\text{c}^2$. We find that $0\leq r_\pm\leq 1$, with the constraint $0\leq r_++r_-\leq 1$. Note also that at $\chi=0$, $u=0$.
Crucially, while Eq.~\eqref{eq:W2by2_eigenvalues} is a well-known result~\cite{Belzig_2005,Nazarov_Blanter_2009},
we here argue that it has not yet been interpreted to its fullest extent. 

As a first step, we point out that for these two eigenvalues, we encounter two distinct topological phases, in accordance with~\cite{Ren_2012}.
The spectrum is either trivial, as depicted in Fig.~\ref{figure_SQD}c,
or topological with a $4\pi$-periodicity, as depicted in Fig.~\ref{figure_SQD}d.
In the latter case, the two eigenvalues $\lambda_{0,1}$ merge into
one single band $\lambda_{\nu=0}$. The occurrence of the topological phase can be easily understood in terms of basic complex analysis, when considering $u(\chi)$, see Eq.~\eqref{eq_z_chi}. For $r_++r_-<1/2$, $u(\chi)$ does not enclose the point $-1$ when drawing it as a complex contour for $\chi$ from $0$ to $2\pi$. Here, the spectrum is trivial, as we can always evaluate the square root in Eq.~\eqref{eq:W2by2_eigenvalues} for all $\chi$ without having to take into account the branch cut. As soon as $r_++r_->1/2$, $u(\chi)$ does enclose the point $-1$, and the eigenspectrum is topological, as the branch cut is unavoidable.

\begin{figure}
\begin{center}\includegraphics[width=0.8\columnwidth]{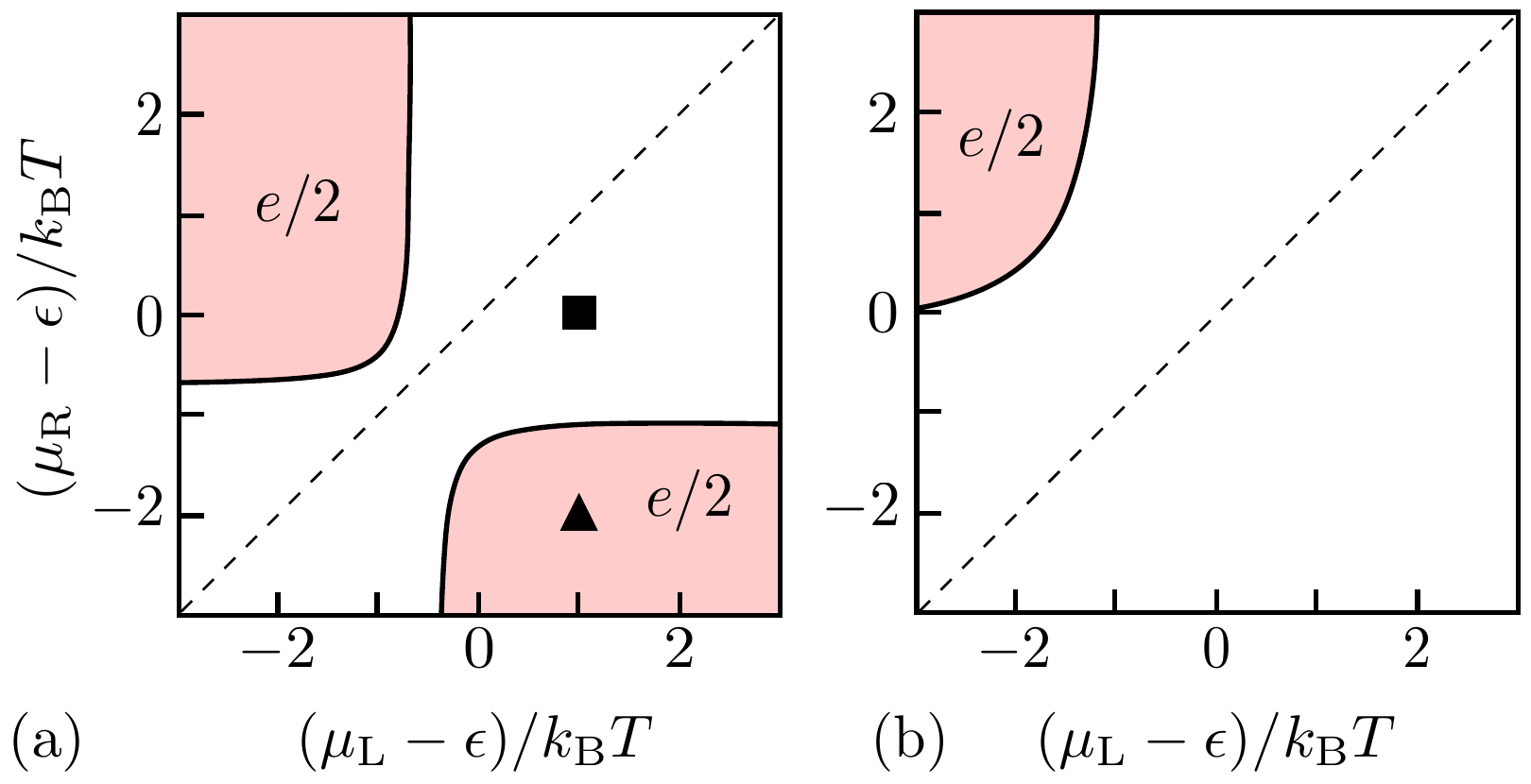}\end{center}

\caption{The different topological phases of transport through the single-level quantum dot, with the corresponding fractional
charge for various parameters, as a function of $\mu_{\text{L}}-\epsilon$
and $\mu_{\text{R}}-\epsilon$. Red areas correspond to
fractional charge $e_{0}^{*}=e/2$. The tunneling asymmetry
is different for the two panels, $\Gamma_{\text{L}}/\Gamma=0.6$ for
(a) and $\Gamma_{\text{L}}/\Gamma=0.8$ (b), respectively.}
\label{figure_SQD_fractional_charge}
\end{figure}

Importantly, as we stated already, the topological phase appears within a connected parameter subspace, see Fig.~\ref{figure_SQD_fractional_charge}a and b, and is thus stable with respect to small variations of the system parameters. The configurations that favour the topological phase are for strong bias, where the transport goes predominantly in one direction, see Fig.~\ref{figure_SQD}b. For the special case of zero bias on the other hand, the braid phase transition is not possible (as also pointed out by~\cite{Ren_2012}, see also Sec.~\apponeBsymmetries). Namely, in equilibrium, the eigenspectrum of $W$ is real for all $\chi$, and is in general gapped, with an ordinary $2\pi$-periodicity. As we will discuss later (in Sec.~\ref{sec_analogy_fJE}), topological Josephson junctions provide an exception, where the transport dynamics is described by an operator with a real spectrum with broken periodicity. This however requires extra conditions, which are satisfied thanks to the presence of the nontrivial superconductors (see also~\apponeChermitianmatrices). Finally, the topological phase in $W$ is inhibited likewise for strongly asymmetric tunnel coupling (even in the presence of a strong bias), because strong coupling asymmetry leads to one of the two tunnelings being the bottle neck process. This results in eigenvalues which are far apart, and do thus not engage in a braid phase transition.

In fact, even though it is hard to predict the topology of the eigenspectrum for a general matrix $W(\chi)$ without explicitly calculating it, the above realizations can serve us to give at least some rudimentary recipes, which are valid beyond the above simple model with only two states. In order to create a system with a nontrivial topology, one should generally avoid being close to equilibrium. As just stated, in equilibrium, the eigenspectrum is real, and is therefore in general gapped, unless additional symmetries are present (see also Sec.~\ref{sec_analogy_fJE}). Secondly, processes occuring with strongly different time scales (such as the strongly asymmetric tunnel coupling mentioned above) give rise to strongly separated eigenvalues that do not partake in a braid phase transition. A third point relates to the fact that in above kernel, we did not include  energy levels and charge states, that are either way above or way below the energy window spanned by the two chemical potentials. These additional states decay in principle with similar rates as the active states $|0,1\rangle$ (at least if the tunneling rates are only weakly energy-dependent), and thus they do not provide a separation of time scales. However, such levels do not contribute to the transport as they are either always empty, or always filled. This fact manifests on the level of the eigenspectrum as eigenvalues that depend only very weakly on $\chi$. Such eigenvalues are inert, and do likewise not partake in a braid phase transition.

Let us now look at the moment generating function in the long measurement time limit $\tau\rightarrow \infty$. Here, only the eigenvalues with the real part closest to zero survive, see Eq.~\eqref{eq:cumulant_eigenvalues}. While for a trivial spectrum this simply means that the moment generating function is given by a single mode, $m(\chi,\infty)=\lim_{\tau\rightarrow \infty}e^{\lambda_0(\chi)\tau}$, for the topological spectrum, $m$ becomes discontinuous, 
\begin{equation}\label{eq_m_disc}
m(\chi,\infty)=\lim_{\tau\rightarrow \infty}\left\{\begin{array}{c}
e^{\lambda_{0}\left(\chi\right)\tau}\quad\text{for }\text{Re}\left[\lambda_{0}\left(\chi\right)\right]>\text{Re}\left[\lambda_{1}\left(\chi\right)\right]\\
e^{\lambda_{1}\left(\chi\right)\tau}\quad\text{for }\text{Re}\left[\lambda_{0}\left(\chi\right)\right]<\text{Re}\left[\lambda_{1}\left(\chi\right)\right]
\end{array}\right.\ .
\end{equation}
As we see, the moment generating function remains $2\pi$-periodic, while the analytic continuation of $m$ around $\chi=0$ leads to a broken periodicity in $\chi$. However, which of the two will be measured? Crucially, the $2\pi$-periodic result will be found by an experimenter having explicitly access to a detector attached to the right reservoir, as introduced in Sec.~\ref{subsec:braiding_in_FCS}. Let us contrast this to a different experimenter, who has access to the cumulants $C_k(\tau\rightarrow\infty)$, e.g., by measuring the expectation values of the current operator $\widehat{I}$ and its higher order correlations for very long times (see Eqs.~\eqref{eq:cumulants_def_2} and~\eqref{eq_cumulant_long_times} in Sec.~\ref{subsec:braiding_in_FCS}). The latter experimenter will not observe the discontinuity. Instead, by reconstructing the cumulant generating function through a Taylor series $c(\chi)=\sum_k (i\chi/e)^kC_k/k!$ and performing the analytic continuation, he must come to the conclusion that a physical fractional charge of $e/2$ is present. He simply performs a Fourier analysis back from $\chi$ to charge space, and he will thus find that the system should be described by a charge operator with non-integer eigenvalues. This latter nonphysical finding is due a subtle issue concerning the order of limits. Namely, the convergence of the cumulant generating function for finite $\chi$ depends on how long one really measures, and to how many orders one sums up the cumulants. The fractional charge would become physical when the limit $\tau\rightarrow\infty$ is taken first. This result may to some extent be considered unproblematic in this limit, since the number of charges transported in this limit is infinite. However, in reality of course, no experiment goes on forever, which means that this result has to be interpreted with care. Thus, we can clearly see now the importance of the explicit presence of a detector: it resolves this issue also in the limit $\tau\rightarrow\infty$, and reinstates charge quantization for all times $\tau$. Namely, it is the $2\pi$-periodicity of $W(\chi)$ which ensures the correct global properties of $m$.

We now relate this result to one example of FCS in strongly correlated transport. In Ref.~\cite{Gutman2010}, the authors study the transport through a 1D nanowire with an interacting region. In the limit where the measurement time $\tau$ is short with respect to the plasmon time-of flight, but long with respect to inverse energy scale given by the quasiparticle distributions, they find for the cumulant generating function,
\begin{equation}
\begin{split}
c(\chi)=\sum_{l=0}^{\infty}\int\frac{d\epsilon}{2\pi}\left\{ \ln\left[1+\left(e^{i\chi e_{l}^{*}}-1\right)n_{\text{L}}\left(\epsilon\right)\right]\right.\\+\left.\ln\left[1+\left(e^{-i\chi e_{l}^{*}}-1\right)n_{\text{R}}\left(\epsilon\right)\right]\right\} \ ,
\end{split}
\end{equation}
where the transport is a sum of left (L) and right movers (R), with their corresponding occupation number $n_\text{L,R}$, which are split into quasiparticles with fractional charges $e^*_l$. Unlike the FCS in a single-level quantum dot, there is an additional sum over energies, since there are no localized levels, and instead a continuum of energy channels is available. Therefore, to make above result more relatable to our quantum dot system, let us consider the contribution of just one energy channel, $\epsilon$. In addition, let us assume a small quasiparticle occupation for this particular energy $n_\alpha(\epsilon) \ll 1$, where
\begin{equation}\label{eq_c_LL}
c_\epsilon(\chi)\approx \sum_{l=0}^{\infty}\left\{ \left(e^{i\chi e_{l}^{*}}-1\right)n_{\text{L}}\left(\epsilon\right)+\left(e^{-i\chi e_{l}^{*}}-1\right)n_{\text{R}}\left(\epsilon\right) \right\}\ ,
\end{equation}
that is, we recover a sum of Poissonian transport events with different charges $e_l^*$. As we will see in a moment, we can find an expression of the same structure for sequential electron transport in a particular limit, see Eq.~\eqref{eq_Poissonian_limit}. 
The moment generating function $m(\chi)=e^{\tau c(\chi)}$, with $c$ given in Eq.~\eqref{eq_c_LL}, is manifestly not $2\pi$-periodic in $\chi$, violating charge quantization, thus presenting a similar issue as the one in sequential tunneling. In Ref.~\cite{Gutman2010}, the authors notice the violation of charge quantization, and assert that is a consequence of neglecting charge fluctuations on the length scale of the Fermi wave length (referring to~\cite{Haldane_1981}). If transport is measured across a sharp interface, the moment generating function is, according to~\cite{Gutman2010}, only correct for $-\pi<\chi<\pi$, and then has to be continued periodically. This forced periodicity comes, just like in our example of conventional sequential electron tunneling, at the expense of discontinuities in $m$ at $\chi=\pm\pi$. As we see, when considering the FCS, the fractional charge is again not an exact property, but a property of the analytic continuation of the moment generating function.
This is a striking analogy of the transport statistics of two extremely different systems. The only difference concerns the values of the fractional charges. Namely, the charges in Luttinger liquid physics, $e_l^*$, can in general be irrational. The actual expressions for $e^*_l$ depend on the details of the system~\cite{Gutman2010}. We refrain from discussing them in detail, as this is not the focus of our work.

At this point, we note that in Ref.~\cite{Nazarov_Blanter_2009}
the idea of fractional charges in sequential tunneling has already been briefly coined in passing. In that reference, the authors consider similarly the FCS of single-electron transistors, and find in a special limit the result
\begin{equation}\label{eq_Poissonian_limit}
c(\chi)=\gamma\left(e^{i\chi/2}-1\right)\ ,
\end{equation}
which looks obviously like the Poissonian statistics of a $e/2$ particle.
We can recover the same result in Eq.~\eqref{eq:W2by2_eigenvalues}, when setting the tunneling rates to $\sigma \Gamma_\text{L}=\Gamma_\text{R}=\gamma$, and the reservoir occupations to $f_\text{L}= 1$ and $f_\text{R}= 0$ (that is, for a very strong bias). In Ref.~\cite{Nazarov_Blanter_2009}, the authors likewise discarded this fractionalization as unphysical, arguing that single electron transistors ``do not chop electrons in half''. However, as we already insisted upon in the introduction, the same is true for strongly correlated systems: electrons can very generally not be split in a literal physical sense, and it is paramount that the moment generating function retain its $2\pi$-periodicity.
Secondly, and very importantly, in Ref.~\cite{Nazarov_Blanter_2009}
the fundamental connection between fractional charges and braid phase transitions has not been considered. The occurrence of fractional charges is \textit{not} some freak coincidence for very special chosen parameters, but is a stable topologically protected property occurring in a large, connected parameter subspace. Moreover, the question is not whether a certain physical effect, be it topological transitions in the FCS, or strong quantum correlations in Luttinger liquids, literally produces fractional charges - neither of them do. The question is, whether the nature of fractional charges occurring in strongly correlated systems \textit{differ} from the ones occurring in conventional sequential electron tunneling. We here conclude, that on the level of the transport statistics in a long measurement time regime, we find no qualitative features to tell them appart.
In the next section, Sec.~\ref{subsec:periodicity_and_charge}, we strongly generalize this statement. In fact, we will provide a concrete argument allowing for the interpretation of \textit{all} the eigenmodes of the system-detector dynamics (i.e., not only the stationary mode $\lambda_0$) in terms of fractional charges.

\subsubsection{Quantum dots in series}

However, before embarking on a general interpretation of higher modes as fractional charges, let us briefly introduce a system which actually produces nontrivial decaying modes. For this purpose, we consider a quantum dot system with more than a single level. For the previous example, we only had only two available states (empty and filled). This naturally limits the number of different topological phases we can observe. For one, we cannot observe a fractional charge of more than 1/2, due to only having 2 states. For another, we cannot hope to see topological transitions that do not involve the stationary state. For the more complicated model presented now we will see a much richer topological phase diagram. Among others, we will find a phase with a $e/3$-charge stationary mode. Secondly, another phase with a trivial stationary mode and a topological decaying mode will emerge, the understanding of which will require the more general interpretation brought forth subsequently, in Sec.~\ref{subsec:periodicity_and_charge}. 

\begin{figure}
\begin{center}\includegraphics[width=1\columnwidth]{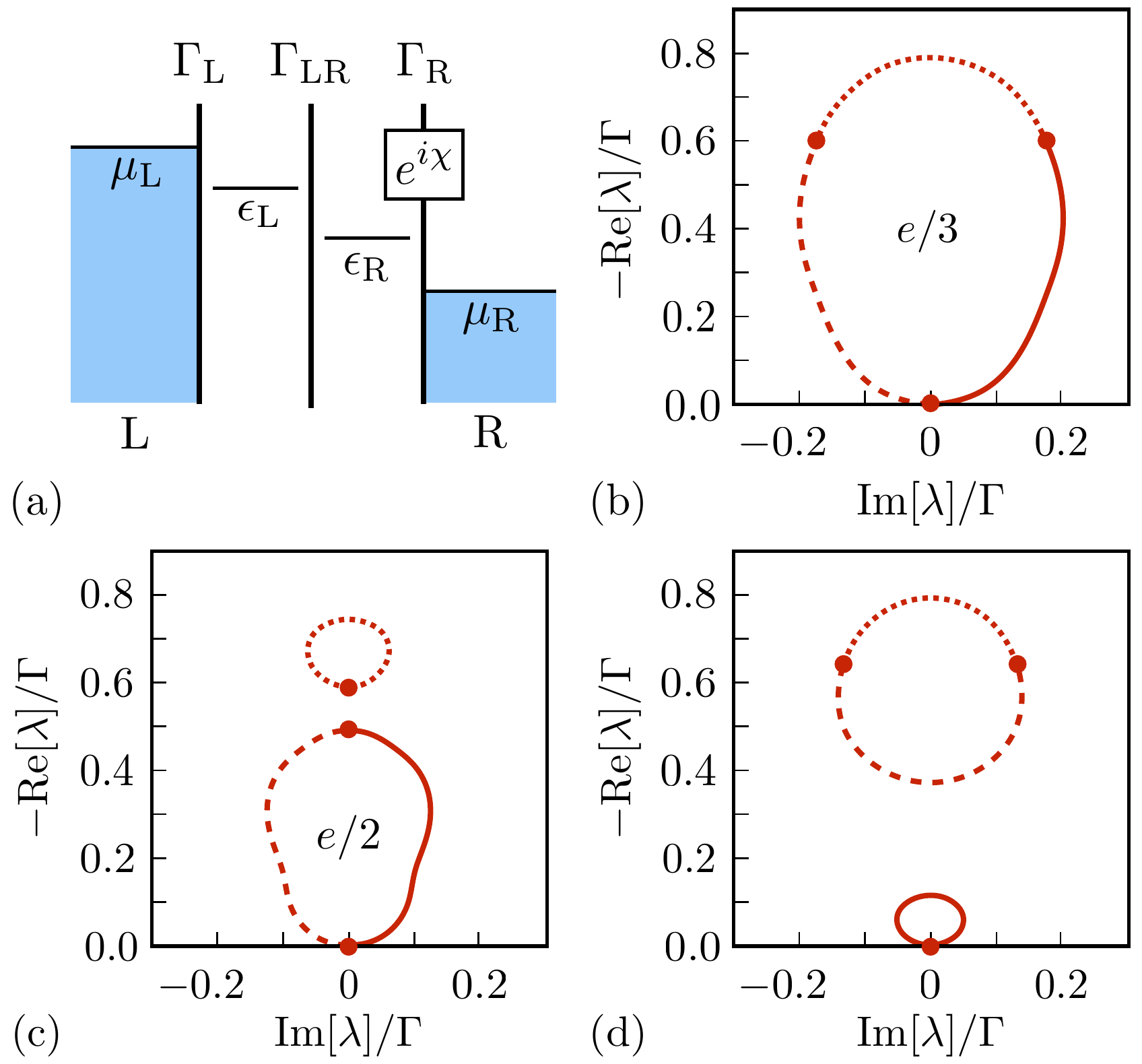}\end{center}

\caption{(a) The model with two quantum dots in series, with the
energies $\epsilon_{\text{L,R}}$, and the sequential electron tunneling
rates $\Gamma_{\text{L},\text{LR},\text{R}}$. The transport statistics
are measured at the interface between the right dot and the right
reservoir. In (b-d) we show the three distinct topological phases that occur
in the system-detector dynamics. In (b) all three eigenvalues merge
into one band with fractional charge $e^{*}=e/3$. Alternatively,
the lowest two eigenvalues can merge into a band with charge $e^{*}=e/2$,
see (c). Finally, the upper two eigenvalues can merge into one band.
The parameters for (b-d) are $\Gamma_{\text{L}}=\Gamma_{\text{R}}=0.3\Gamma$
and $\Gamma_{\text{LR}}=0.4\Gamma$, with $\Gamma=\Gamma_{\text{L}}+\Gamma_{\text{R}}+\Gamma_{\text{LR}}$
as well as $\mu_{\text{L}}=2k_{\text{B}}T$, $\mu_{\text{R}}=-2k_{\text{B}}T$
and $\epsilon_{\text{R}}=k_{\text{B}}T$. The remaining parameter
is $\epsilon_{\text{L}}/k_{\text{B}}T=\left\{ 0,1.5,3.2\right\} $
for (c), (b), and (d), respectively. }
\label{figure_DQD}
\end{figure}

We consider two quantum
dots in series, see Fig.~\ref{figure_DQD}a. We assume that again the
individual dots have very high charging energies, and that the gates
are tuned such that only the transitions between the three charge
states $\left|00\right\rangle $, $\left|10\right\rangle $, and $\left|01\right\rangle $
(i.e., zero extra charges, or a left or right extra charge) are available.
As for the charge exchange between the two dots, we focus on a regime
where the interdot transport is dominated by an inelastic, incoherent
process. For the full Hamiltonian description of this model, consult Sec.~\appdoublelevelQD.

We may thus capture the dynamics again by means of a fully diagonal
density matrix, with the vector of probabilities $\left|P\right)=\left(P_{00},P_{10},P_{01}\right)$,
and through the kernel
\begin{equation}
W\left(\chi\right)=W_{\text{L}}+W_{\text{LR}}+e^{i\chi\widehat{n}_{R}}W_{\text{R}}e^{-i\chi\widehat{n}_{R}}\ ,\label{eq:W_3by3-1}
\end{equation}
with $\widehat{n}_{\text{L}}=\text{diag}\left[\left(0,1,0\right)\right]$
and $\widehat{n}_{\text{R}}=\text{diag}\left[\left(0,0,1\right)\right]$.
The matrices 
\begin{align*}
W_{\text{\text{L}}} & =\left(\begin{array}{ccc}
-\sigma\Gamma_{\text{L}}f_{\text{L}} & \Gamma_{\text{L}}\left[1-f_{\text{L}}\right] & 0\\
\sigma\Gamma_{\text{L}}f_{\text{L}} & -\Gamma_{\text{L}}\left[1-f_{\text{L}}\right] & 0\\
0 & 0 & 0
\end{array}\right)\\
W_{\text{\text{R}}} & =\left(\begin{array}{ccc}
-\sigma\Gamma_{\text{R}}f_{\text{R}} & 0 & \Gamma_{\text{R}}\left[1-f_{\text{R}}\right]\\
0 & 0 & 0\\
\sigma\Gamma_{\text{R}}f_{\text{R}} & 0 & -\Gamma_{\text{R}}\left[1-f_{\text{R}}\right]
\end{array}\right)\ ,
\end{align*}
account for the sequential tunneling to and from the contacts, where
the Fermi functions are now $f_{\text{L,R}}=1/\left(1+e^{\beta\left[\epsilon_{\text{L,R}}-\mu_{\text{L,R}}\right]}\right)$.
The matrix
\begin{equation}
W_{\text{LR}}=\left(\begin{array}{ccc}
0 & 0 & 0\\
0 & -\gamma_{\text{R}\rightarrow\text{L}} & \gamma_{\text{L}\rightarrow\text{R}}\\
0 & \gamma_{\text{R}\rightarrow\text{L}} & -\gamma_{\text{L}\rightarrow\text{R}}
\end{array}\right)\ ,\label{eq:W_LR_inelastic}
\end{equation}
accounts for the tunneling between the two dots. The specific shape
of $\gamma_{\text{L}\rightarrow\text{R}},\gamma_{\text{R}\rightarrow\text{L}}$
will depend on the details of the mechanism mediating the interdot tunneling
process, e.g., through electron-phonon mediated tunneling \cite{Fujisawa_1998,Hu_2005}.
For the sake of simplicity, we assume that they are of the form $\gamma_{\text{L}\rightarrow \text{R}}=\Gamma_{\text{LR}}f\left(\epsilon_{\text{L}}-\epsilon_{\text{R}}\right)$,
$\gamma_{\text{L}\rightarrow\text{R}}=\Gamma_{\text{LR}}f\left(\epsilon_{\text{R}}-\epsilon_{\text{L}}\right)$,
resulting in temperature dependent rates, where for large detuning,
$\left|\epsilon_{\text{L}}-\epsilon_{\text{R}}\right|\gg k_{\text{B}}T$,
there remains simply a constant rate towards the lower lying level
(neglecting any energy-dependence of $\Gamma_{\text{LR}}$). For one, we stress
that since we are merely interested in studying the general topological
properties, it is not necessary to take into account more details. Secondly, we point out that the same simple energy dependence for the inelastic rate was successfully used to describe inelastic tunneling in a two-atom electron pump~\cite{Roche_2013}. 

As for the topology of the kernel, we find the following. Apart from
the trivial phase with three independent bands, the system with two
quantum dots (islands) in series gives rise to $3$ distinct topologically
nontrivial phases. They are depicted in Fig.~\ref{figure_DQD}b to
d. In Fig.~\ref{figure_DQD}b all three eigenvalues form a single
band, while in Fig.~\ref{figure_DQD}c, the two lower eigenvalues
form the same topological stationary mode as in the previously discussed
single quantum dot. We can also observe a topological phase, where
the stationary mode is trivial, while the two decaying modes are braided,
see Fig.~\ref{figure_DQD}d.

In close analogy to the above discussion, the first two topological phases
(b and c), again gives rise to fractional charges in the stationary mode. In particular, for the first topological
phase (b), there may occur a
charge $e/3$. 

The third topological phase (d) will however provide trivial statistics
in the here considered limit of long measuring times. In the following
section, we will analyze the interplay between system and transport
detector from a very different vantage point, and for finite $\tau$. This allows for a much
more general interpretation of the topological phases
as fractional charges, including the decaying modes, as in Fig.~\ref{figure_DQD}d.

\subsection{Braid phase transition and detector resolution\label{subsec:periodicity_and_charge}}

As we have seen, charge quantization is important to correctly describe the global properties of the moment generating function, and that fractional charges can only be well-defined in the infinite measurement time limit as the analytic continuation of $m$ in $\chi$. Now, we want to generalize our findings to finite measurement times.

\begin{figure}
\begin{center}\includegraphics[width=0.9\columnwidth]{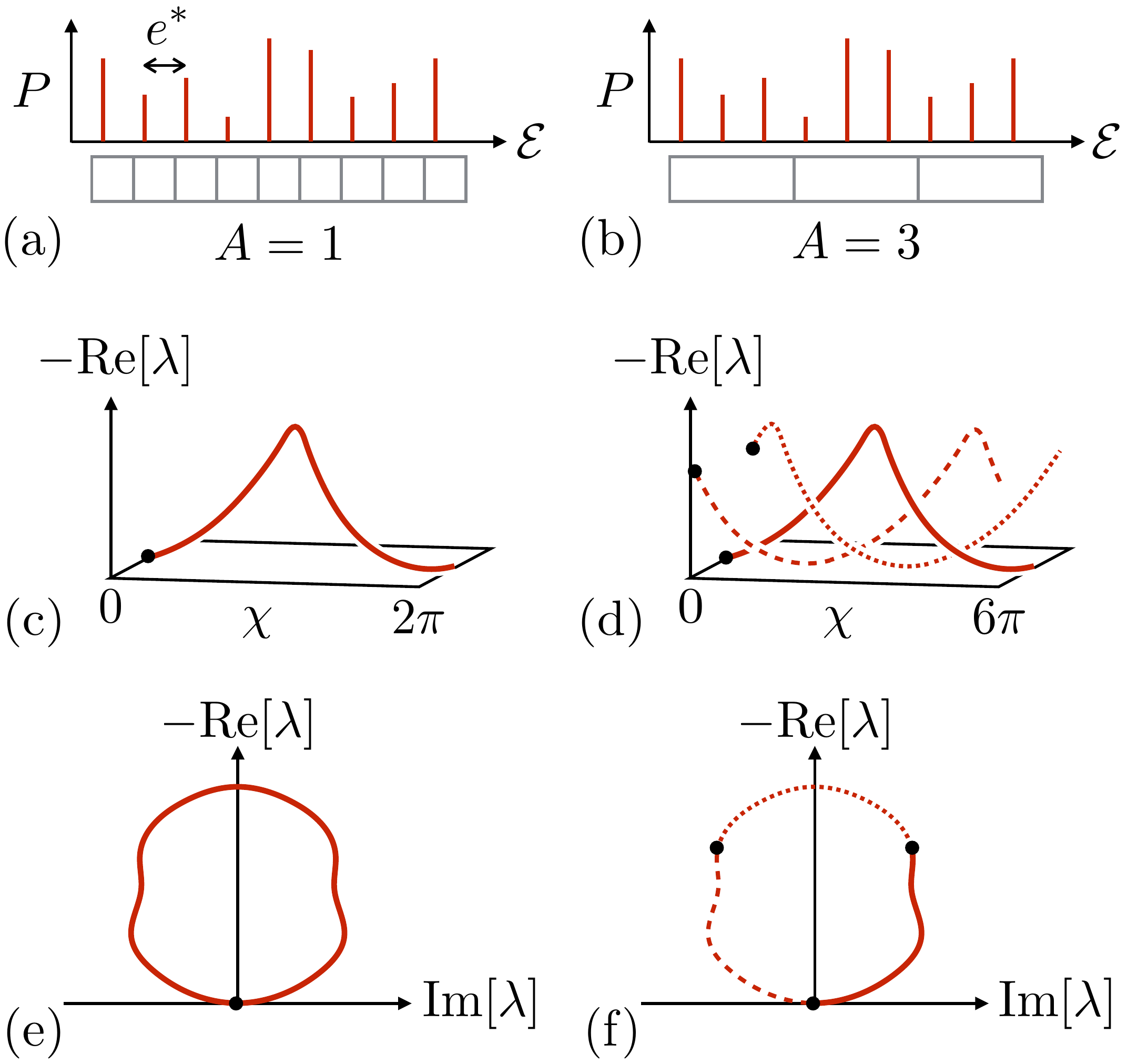}\end{center}

\caption{The concept of detectors with insufficient resolution. In (a) we show the probability density $p$ as a function of $\mathcal{E}$
for an ideal detector, measuring the transport statistics of a particle
with charge $e^{*}$. Below the $\mathcal{E}$-axis, we indicate a
discretized (pixellated) detector with limited resolution, here exactly the resolution
of $e^{*}$, $A=1$, hence no information is lost through the discretization. (b) Here a discretized
detector with resolution $A=3$ is shown. The charge transport can
no longer be resolved exactly. (c) A generic depiction of a complex
eigenmode $\lambda$ as a function of $\chi$, appearing in the detector
dynamics. (d) The eigenmodes $\widetilde{\lambda}_{a}$ as a function
of $\chi$ appearing in the time evolution of the detector state with
insufficient resolution, $A=3$. We see that the insufficient resolution
results in the appearance of shifted copies of $\lambda\left(\chi\right)$,
resulting in the same type of braid topology as in Fig.~\ref{fig_generic_spectrum}.
(e) and (f). The spectra of (c) and (d), respectively, are projected
onto the complex plane, for $0\leq\chi<2\pi$. The black dots indicate
the eigenvalues at $\chi=0$. \label{fig_detector_resolution}}
\end{figure}

In order to do so, we propose an analogy between charge quantization and the notion of a detector with a minimal resolution limit. To begin, let us consider a very generic, out-of-equilibrium transport situation with two reservoirs, exchanging particles with a given charge. For the moment, we now explicitly deviate from the charge quantization requirement, such that the eigenvalues of the charge operators in the reservoirs do not necessarily need to be a multiple integer of $e$. Let us first restrict our discussion to the case, where only a single type of charge is exchanged, called 
$e^{*}$. We supplement such a system with an idealized charge detector,
measuring the transport into one of the reservoirs, with the following
two properties. First, it has a continuous degree of freedom, here
denoted as $\mathcal{E}$, corresponding to an infinitely precise
charge resolution. Second, it is ideally coupled to the system, such that
whenever there is a transport event passing a certain charge $e^{*}$
into or out of the reservoir, the detector state receives a kick,
such that $\mathcal{E}\rightarrow\mathcal{E}\pm e^{*}$.
We may describe the corresponding detector state with a density matrix
$\rho\left(\mathcal{E},\mathcal{E}'\right)$, providing the probability
density $P\left(\mathcal{E}\right)\equiv\rho\left(\mathcal{E},\mathcal{E}\right)$. Let us assume that at the inital time $t=0$ the detector state is
reset, $\mathcal{E}=0$ (i.e., no charge transfer is registered before
the measurement starts), and only one type of charge carrier is involved
in the transport.
Then, the probability density at the measurement
time $t=\tau$ (where the detector state is projected onto the charge eigenbasis) must be of the form (see also Fig.~\ref{fig_detector_resolution}a)
\begin{equation}
P\left(\mathcal{E},\tau\right)=\sum_{N}P\left(N,\tau\right)\delta\left(\mathcal{E}-e^{*}N\right)\ .
\end{equation}
Through a Fourier transform, we receive $P\left(\chi,\tau\right)=\int d\mathcal{E}e^{i\frac{\chi}{e}\mathcal{E}}P\left(\mathcal{E},\tau\right)=\sum_{N}e^{i\frac{e^{*}}{e}\chi N}P\left(N,\tau\right)$. We see now that
the discreteness of charge $e^*$ corresponds to a periodicity of $P\left(\chi,\tau\right)$
in $\chi$, here however with period $2\pi e/e^{*}$.

Now we render the detector nonideal, by introducing a finite resolution
in form of ``pixels'', which may have a size larger than $e^*$. Based on the previous discussion, this pixellation can be motivated by multiple  different reasons. As the FCS of sequential tunneling are concerned, we note that in the models we considered so far, the detector is locally attached to the interface
of one reservoir (the right one, in the above examples), and is by construction
``blind'' to all charge transfer processes that do not change the
charge in said reservoir. These include for instance the tunneling into the quantum system from the left reservoir, or the tunneling processes within the quantum system (e.g. the interdot tunneling in the double quantum dot). In a very similar way, we can think of the pixellated detector as ``blind'', since it does not register stochastic events that occur within the same pixel.

We can however motivate the pixellation also in a similar spirit to the FCS of Luttinger liquids presented in Ref.~\cite{Gutman2010}. Namely, a detector with pixel size $e$ can be understood as an ad hoc recipe to reintroduce integer charge quantization. This treatment is in fact very similar to the ``by hand'' reintroduction of charge quantization that the authors of Ref.~\cite{Gutman2010} propose themselves. We will comment in more detail on the similarities and differences to~\cite{Gutman2010} below.

Finally, a third motivation for the detector pixellation will be exploited in Sec.~\ref{subsec:beyond_resolution_limit}, where we consider a possible application of topological detector dynamics when the detector that has a limited resolution rather due to a deficiency than due to a fundamental limit. 

We define the nonideal detector probability as $\widetilde{P}\left(M\right)=\int_{\mathcal{E}_{0}+\mathcal{A}M}^{\mathcal{E}_{0}+\mathcal{A}M+\mathcal{A}}d\mathcal{E}P\left(\mathcal{E}\right)$,
such that all measurement outcomes in the interval $\mathcal{E}_{0}+\mathcal{A}M\leq\mathcal{E}\leq\mathcal{E}_{0}+\mathcal{A}M+\mathcal{A}$
get projected onto a discrete detector state $M\in\mathbb{Z}$, where
$\mathcal{A}$ is the resolution of the nonideal detector, and $\mathcal{E}_{0}$
is an arbitrary and irrelevant shift (for simplicity, we require $0<\mathcal{E}_{0}<\mathcal{A}$).
For our purposes, it is sufficient to consider a resolution $\mathcal{A}$
which is an integer multiple of $e^{*}$, $\mathcal{A}=Ae^{*}$, $A\in\mathbb{N}$
(see Figs.~\ref{fig_detector_resolution}a and b). Then, we find that
the probability distribution of the discretized detector is
\begin{equation}
\widetilde{P}\left(M,\tau\right)=\sum_{a=0}^{A-1}P\left(AM+a,\tau\right)\ .
\end{equation}
Thus, the projected state $M$ sums up the probabilites of being in the states $N$ that are within
an interval from $N=AM$ to $N=AM+A-1$. We likewise define a Fourier transform
$\widetilde{P}\left(\chi,\tau\right)=\sum_{M}e^{iM\chi}\widetilde{P}\left(M,\tau\right)$.
We note that for $A=1$, $\widetilde{P}\left(\chi,\tau\right)=P\left(\chi,\tau\right)$,
which means that if the resolution of the discrete detector matches
$e^{*}$, we have not actually lost any information, as can be expected.

Now we consider the nonideal case, $A>1$.
For the sake of simplicity, let us assume that the dynamics of the original detector
state can be described through a single mode, $P\left(\chi,\tau\right)\rightarrow e^{\lambda\left(\chi\right)\tau}$
(for a generic example, see Figs.~\ref{fig_detector_resolution}c
and e). We note that while the description of the detector dynamics in terms of eigenmodes is obviously motivated by our considerations of sequential tunneling, but my no means restricted to it, see, e.g., the result of Ref.~\cite{Gutman2010}, and references therein. Requiring that $\lambda\left(\chi\right)$ be analytic, one
can show in a straightforward manner that for the nonideal detector
\begin{equation}
\widetilde{P}\left(\chi,\tau\right)=\sum_{a=0}^{A-1}\widetilde{\alpha}_{a}\left(\chi\right)e^{\widetilde{\lambda}_{a}\left(\chi\right)\tau}\ ,\label{eq:cumulant_copies}
\end{equation}
with $\widetilde{\alpha}_{a}\left(\chi\right)=\sum_{a'=0}^{A-1}e^{-i\left[\chi/A+2\pi a/A\right]a'}/A$
and
\begin{equation}
\widetilde{\lambda}_{a}\left(\chi\right)=\lambda\left(\chi/A+2\pi a/A\right)\ .
\end{equation}
We see that the time-evolution for the discretized detector with insufficient
resolution gives rise to a superposition of modes $\widetilde{\lambda}_{a}\left(\chi\right)$,
which are simply shifted copies of the original $\lambda\left(\chi\right)$,
stretched to a new periodicity $2\pi A$, and braided (see Figs.~\ref{fig_detector_resolution}d and f). And the same is true for the coefficients $\alpha_a$. Thus, the new moment generating function provides eigenmodes with the exact same properties - broken periodicity
and redundancy - as the topological subbands introduced in Sec.~\ref{subsec:braiding_in_FCS}.

Crucially, if we now return to the model systems with sequential electron
tunneling, Sec.~\ref{subsec:explicit_examples}, we find that the
detector statistics exhibit a topological eigenspectrum with broken periodicity,
even though the detector is ideal in the sense that it measures exactly
the electrons that physically tunnel into a reservoir. The braid phase
transition does not come from a coarse-grained measurement, but stems
from the extra degrees of freedom of the quantum system to which the
detector is insensitive. 
Here, we study rather generically the dynamics of a detector with
a finite resolution that does not necessarily match the unit of a
certain discrete process - and we make the striking observation, that also
here the same topological eigenspectrum with broken periodicity occurs: the
broken periodicity directly relates to the mismatch between the detector
resolution ($e$) and the sub-pixel size of a discrete process ($e^*$). Therefore, we conclude
that if the dynamics of the detector state is the only accessible
information, then the nontrivial transport statistics with periodicity
$p$ of regular electrons are \textit{topologically indistinguishable}
from trivial transport of fractional charges $e/p$, measured with
a detector with resolution $e$.

Importantly, we can easily generalize the above discussion to the case where
the detector dynamics $P\left(\chi,\tau\right)$ contains more than
just a single mode. Consequently, the equivalence between braid phase transitions
and finite detector resolution can be easily extended to the
decaying modes. For this purpose, we start with the cumulant generating function as given
in Eq.~(\ref{eq:cumulant_eigenvalues}). Taking then into account
the redundancy due to broken periodicity, we can reexpress the moment
generating function in terms of the reduced band index $\nu$, such
that
\begin{align}
m\left(\chi,\tau\right) & =\sum_{n}\alpha_{n}\left(\chi\right)e^{\lambda_{n}\left(\chi\right)\tau}\nonumber \\
 & =\sum_{\nu}\sum_{a=0}^{p_{\nu}}\widetilde{\alpha}_{\nu}\left(\left[\chi+2\pi a\right]/p_{\nu}\right)e^{\widetilde{\lambda}_{\nu}\left(\left[\chi+2\pi a\right]/p_{\nu}\right)\tau}\ ,\label{eq:analogy_general}
\end{align}
where
\begin{equation}\lambda_{n=0}\left(\chi\right)=\lambda_{n=1}\left(\chi-2\pi\right)=\lambda_{\nu=0}\left(\chi\right)=\widetilde{\lambda}_{\nu=0}\left(\chi/2\right)\ ,
\end{equation}
and
\begin{equation}
\begin{split}\lambda_{n=2}\left(\chi\right)=\lambda_{n=3}\left(\chi-2\pi\right)=\lambda_{n=4}\left(\chi-4\pi\right)\\=\lambda_{\nu=1}\left(\chi\right)=\widetilde{\lambda}_{1=0}\left(\chi/3\right)\ ,
\end{split}
\end{equation}
and likewise for the coefficients $\alpha$.
 
\begin{figure}
\includegraphics[width=0.95\columnwidth]{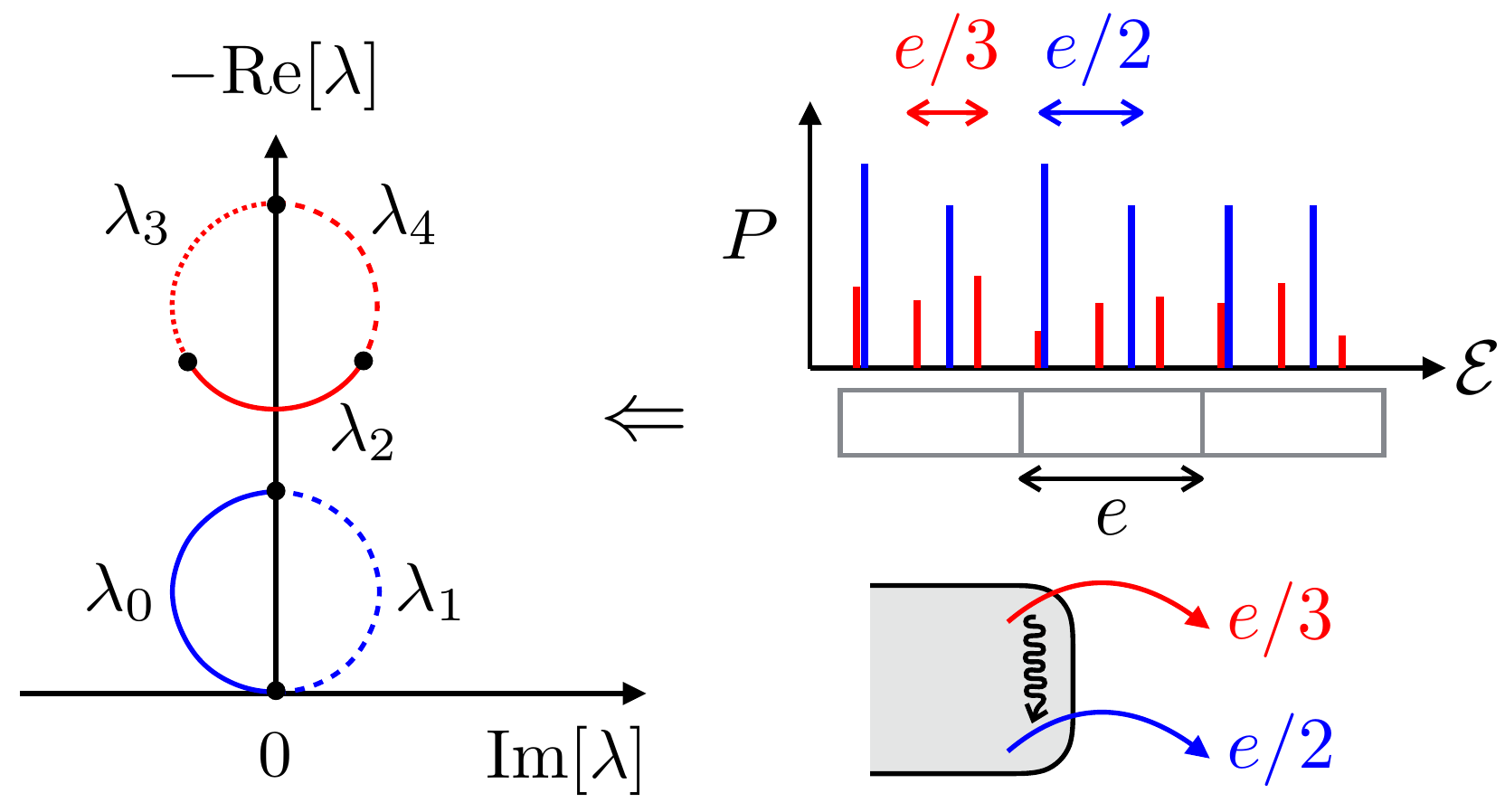}

\caption{The analogy between a topological eigenspectrum with
broken periodicity, and the transport of (fractional) charges measured
with a pixellated detector. On the left hand, we show again the generic
example of the eigenspectrum of Fig.~\ref{fig_generic_spectrum}.
On the upper right hand, we show a pixellated detector setup that
gives rise to the spectrum on the left side. Thus, the spectrum on
the left is equivalent to the transport of $e/2$ and $e/3$
charges, measured with a detector that has a minimum resolution of
the elementary charge $e$. The fact that one of the bands (the red
band with charge $e/3$) is decaying, can be interpreted
in simplified terms as an emitter with two possible emission states
(see lower right side): the emission state of the $e/3$ band is unstable,
eventually decaying to the $e/2$ emission state.\label{fig_resolution_analogy}}

\end{figure}

We thus conclude, that the above dynamics
is equivalent to the coexisting transport of $e/2$ and $e/3$ charges,
which are measured with a detector that can only resolve the transport
in units of the elementary charge $e$, see Fig.~\ref{fig_resolution_analogy}.
We can think of this transport in terms of the simple picture of an
emitter that is either in a state where it emits $e/2$-charges, or
in a state that emits $e/3$-charges. The fact that the $e/3$-band
is decaying can then in addition be regarded as the $e/3$ emission
state being unstable, i.e., when the emitter is in this state, it
relaxes to the stable $e/2$ emission state with a given rate.
Along these lines, we are now also able to understand the
third topological phase for the double quantum dot, depicted in Fig.~\ref{figure_DQD}d.
Namely, it corresponds to a stable emission state of regular electrons, and an unstable emission state with a fractional charge $e/2$.

To state the above in generalized terms, we find that each band $\nu$,
with periodicity $p_{\nu}$, contributes to the transport with the fractional charge
\begin{equation}
e_{\nu}^{*}=\frac{e}{p_{\nu}}\ .
\end{equation}
This is a central result of this paper, establishing the initial claim
depicted in Fig.~\ref{fig_generic_spectrum}.

While we already refer to the eigenmodes as contributions with a fractional charge, we would like to concretely ask the question, to which extent can we think of these eigenmodes as fractionally charged \textit{quasiparticles} in a similar sense to the excitations appearing in Luttinger liquids or the FQHE. We can provide two arguments in favour of such an interpretation.
First, through above procedure we can draw a straightforward correspondance from topologically nontrivial FCS in sequential tunneling to an auxiliary system which contains excitations described by a charge operator with fractional eigenvalues, $e_{\nu}^{*}$, which is afterwards requantized by invoking the notion of a minimal detector resolution of size $e$. Of course, these fractionally charged quasiparticles are thus fictitious, because they have to be supplemented by a detector with integer resolution in order to be physical. But, as we have already stressed on several occasions, we do not believe that this notion of fractional charge is neither more nor less fictitious than the one of quasiparticles in strongly correlated systems, since charge quantization in units of $e$ is fundamental for electronic systems.

This leads us to the second argument. Namely, we want to point out a possible relation of the above introduced detector resolution to the reintroduction of integer charge quantization in Luttinger liquid theory, as it was proposed in Ref.~\cite{Gutman2010}. To briefly repeat, we pointed out in Sec.~\ref{subsec:explicit_examples} that the transport statistics in Luttinger liquids gives rise to a moment generating function with broken periodicity, thus explicitly violating charge quantization.
Gutman et al.~\cite{Gutman2010} propose to reintroduce charge quantization by simply accepting the values of the moment generating function only for $\chi$ from $-\pi$ to $\pi$, and continuing the function periodically beyond that interval. We could alternatively 
reintroduce charge quantization in their result through adding a finite detector resolution of $e$, along the lines presented above. Note that even though the procedure was motivated for completely classical dynamics, the generalization to quantum transport is possible, because the system is allowed to evolve coherently until the projective measurement at time $\tau$, where we perform the pixellation. In fact, this procedure provides a similar result as the one in Ref.~\cite{Gutman2010} except that it does not simply discard the values of the moment generating function for $|\chi|>\pi$ beyond $\pi$.
Instead, due to the copying of eigenvalues described above, these values would be folded back onto the interval between $-\pi$ and $\pi$, as decaying modes, such that the interval $-\pi<\chi<\pi$ can be considered as a Brillouin zone (again with $\chi$ being a detector momentum) with a unit cell size given by $e$.
For finite times, the two procedures provide a moment generating function with the same lowest mode, but where our procedure provides additional extra modes which decay exponentially with time. For long times, the two results eventually converge, where the decaying bands no longer contribute. We emphasize that at this point, both methods to enforce a $2\pi$-periodicity are completely ad hoc, and we cannot say which of the two, if any, can be formally justified, as such an effort would go way beyond this work. As far as Luttinger liquid theory is concerned, generalizations have been developed~\cite{Imambekov2012}, however the issue of integer charge quantization has to our best knowledge not been discussed~\cite{Schmidt_PC_2019}. We emphasize though that later in this work, in Sec.~\ref{sec_analogy_fJE}, we consider the special case of superconducting transport across Josephson junctions, where the $2\pi$-periodicity, and thus the charge quantization in terms of multiples of $e$ (that is, actually $2e$, since we will consider supercurrents), can be justified on formal grounds, even in the presence of strong correlations, where we can thus formulate an even closer analogy.

Importantly, the ramifications of our result may well be, that transport statistics generally turn out to be an ill-suited observable to unequivocally identify exotic quasiparticles, such as Laughlin quasiparticles, or parafermions. At the very least in the here considered low-frequency regime of FCS, based on the topological arguments presented above, we do not see any qualitative feature which would distinguish them from the fractional effect in sequential tunneling. More research will be required to definitively answer this question, e.g., by a better understanding of charge quantization in Luttinger liquid theory, or by comparing topological features in the high-frequency transport statistics. 
 
Finally, we stress that surprisingly, the pixellation
of the detector as described above does not actually lead to a loss
of information about the detector dynamics: the eigenvalues $\lambda_{n}\left(\chi\right)$
are still exactly observable even when the detector has insufficient
resolution. We will provide a concrete example
within the context of quantum transport, where this principle can
be exploited (see Sec.~\ref{subsec:beyond_resolution_limit}).  Finally, we note that we have so far included only charges that are rational fractions of $e$. The more general case when $e$ and $e^*$ are incommensurable, may be interesting for several reasons, such as to strengthen the analogy to Luttinger liquids (where quasiparticles with irrational fractions are possible), or also in the context of measuring beyond the detector resolution.

\section{Nonequilibrium geometric
phases and analogy to fractional Josephson-effect\label{sec:fractional_josephson_effect}}

In above sections, we have emphasized the nontrivial relation between fractional charges, related to the periodicity of the eigenspectrum, and integer charge quantization, embedded in the global periodicity of the kernel $W$. For the particular example of sequential electron tunneling through quantum dots, we have seen that the braid phase transition arises from transport being measured locally, at a sharp interface, missing certain tunneling events. We were able to relate this principle to a general notion of a detector with a finite resolution (given by the integer charge) and a discrete process with a sub-resolution size (the fractional charge). We will show in this section that there arises a nontrivially quantized geometric phase, due to the mismatch between the two. Namely, this geometric phase contains additional information beyond the topology of the eigenvalues: namely, it is nontrivial when the periodicity of the eigenspectrum is \textit{different} from the periodicity of the kernel $W$. The geometric phase we consider here, arises from a parallel vector transport of the eigenvectors of $W(\chi)$ along $\chi$, giving rise to connections of the form $\sim\left(\nu\right|\partial_{\chi}\left|\nu\right)$. Considering the number of transported electrons $N$ and the counting field $\chi$ as conjugate quantities, we can think of such a phase as a generalized version of a 1D Zak-Phase, for generally non-Hermitian matrices $W$. In a second part, we will use this result to demonstrate that the fractional effect in sequential electron transport provides a profound classical analogy to the fractional Josephson effect emerging in topological superconducting junctions.

We note that while topological invariants defined through geometric phases in parallel
vector transport are well-established in the context of \textit{closed}
topological quantum systems, their
generalization to \textit{open} quantum systems is still an actively considered, open problem~\cite{Sinitsyn_2007,Sinitsyn07PRL,Sinitsyn_2009,Ren2010Apr,Diehl_2011,Bardyn_2013,Budich_2015,Pluecker_2016,Pluecker_2017,Leykam_2017,Kunst_2018,Bardyn_2018,Edvardsson_2019}. Our work contributes to the search for appropriate geometric phases in open quantum systems, by proposing a phase defined through the detector degrees of freedom.

\subsection{Relation between geometric phase and fractional charge}\label{sec_geometric_phase}

In order to define the geometric phase, we perform a parallel vector transport of $|n(\chi))$ along $\chi$ (we will explain in Sec.~\ref{subsec:Measuring_geometric_phase}, how such a vector transport can be measured physically). We take the projector $|n(\chi))(n(\chi)|$ at a certain $\chi_0$ and repeatedly apply it to itself for increasing $\chi_j=j\Delta\chi+\chi_0$, for the integer $j$ going from $0$ to $J$. Going to the continuum limit $\Delta\chi\rightarrow 0$, $J\rightarrow \infty$ (while keeping $\chi_J=J\Delta\chi$ finite) and taking the trace of the resulting matrix object, we receive a phase factor times an overlap function
\begin{equation}\label{eq_geom_def}
\begin{split}
\lim_{\Delta\chi\rightarrow 0}\text{tr}\left[\prod_{j=0}^J |n(\chi_j))(n(\chi_j)|\right]=\\e^{-\int_{\chi_0}^{\chi_J}d\chi (n(\chi)|\partial_\chi|n(\chi))}(n(\chi_0)|n(\chi_J))\ .
\end{split}
\end{equation}
This quantity is manifestly gauge invariant. If we want to completely isolate the phase factor, and get rid of the overlap function $(n(\chi_0)|n(\chi_J))$,
we need to continue our path in $\chi$ until we end up in the same eigenvector as the initial one. For a generally topological eigenspectrum, this means that we may need to advance $\chi$ by more than just $2\pi$. In fact (and as already stated in Sec.~~\ref{subsec:braiding_in_FCS}) due to the $2\pi$-periodic $W(\chi)$, both the eigenvalues and eigenvectors have the same redundancy, such that we may likewise switch to the reduced band index $\nu$ for the geometric phases. Choosing a $2\pi p_\nu$-periodic gauge for the eigenvectors, and setting the initial $\chi_0=0$ (without loss of generality) we may relate Eq.~\eqref{eq_geom_def} to the phase
\begin{equation}
\mathcal{B}_\nu=e^{-\int_0^{2\pi p_\nu}(\nu(\chi)|\partial_\chi|\nu(\chi))}\ .
\end{equation}
First of all, let us verify that this actually corresponds to a real phase, that is, $i\int_0^{2\pi p_\nu}(\nu(\chi)|\partial_\chi|\nu(\chi))\in\mathbb{R}$. In fact this is ensured by the symmetry $W^*(\chi)=W(-\chi)$, which originates from the fact that all transition rates in $W$ are real. As long as we follow bands with $\lambda_\nu^* (\chi)=\lambda_\nu^*(-\chi)$, it follows likewise that the geometric phase is real~\footnote{There is an exception: in a general matrix with the property $W^*(\chi)=W(-\chi)$ there can in principle occur pairs of eigenvalues with $\lambda_{\nu_1}^*(\chi)=\lambda_{\nu_2}(-\chi)$. Here, the individual geometric phases of either $\nu_1$ or $\nu_2$ are not guaranteed to be real. The product of the geometric phases however still is. In the example systems where we study the geometric phase explicitly, we do not encounter such a case, which is why we do not discuss this in detail.}.

Now we investigate the connection between the geometric phases and the braid topology. Crucially, we can show that while the geometric phases for each individual band are in general not quantized, their product is. Namely, we can demonstrate that for a \textit{general} kernel $W\left(\chi\right)$
\begin{align}
\mathcal{Z}&=\prod_{\nu}\mathcal{B}_\nu=\prod_{\nu}\left(-1\right)^{p_{\nu}-1}=\prod_{\nu}\left(-1\right)^{e/e_{\nu}^{*}-1}\ .\label{eq:znumber}
\end{align}
Due to $e/e_{\nu}^*=p_{\nu}\in\mathbb{N}$, $\mathcal{Z}$ is manifestly
$=\pm1$. This product of geometric phases thus directly depends on the ratios of the periodicity of the eigenspectrum \textit{with respect} to the periodicity of $W(\chi)$ itself, indicating the mismatch of the fractional charges of the quasiparticles with the integer charge quantization (i.e., the detector resolution). Importantly, the eigenvalues themselves do not contain the information of the periodicity of $W(\chi)$, which is why $\mathcal{Z}$ carries additional information specific about the detector. This is a further main result of this
paper. Let us stress in addition the importance of taking into account the redundancy of bands. Had we defined $\mathcal{Z}$ through the product of $n$ instead of $\nu$, it would always be trivially $+1$, independent of the topology.

The full derivation of Eq.~\eqref{eq:znumber} is detailed Sec.~\apponeDZfromgeometricphase. We here briefly recapitulate the main steps and assumptions to demonstrate the quantization of $\mathcal{Z}$.
The chief requirement for deriving Eq.~(\ref{eq:znumber})
is that $W\left(\chi\right)$ can be decomposed into the form $W\left(\chi\right)=\sum_{N}e^{i\chi N}W_{N}$,
that is, a kernel that describes the statistics of some discrete process. The proof then involves a generalization to complex counting fields $\rightarrow\chi\in \mathbb{C}$. As we already stated in Sec.~\ref{subsec:braiding_in_FCS}, in this complex space, we find that $W(\chi)$ exhibits isolated points, so-called exceptional points, where $W$ has no longer well-defined left and right eigenvectors. To each exceptional point one can associate a braid generator (along the lines of Artins braid theory~\cite{Artin_1947}). As far as the geometric phase is concerned, the integrals in Eq.~(\ref{eq:znumber}) can be reduced into a product of phase contributions which enclose the individual exceptional points. We then find that each exceptional point contributes a phase $e^{i\pi}=-1$ to the total phase. This particular step is in accordance with a similar result found in Ref.~\cite{Leykam_2017}. The final step then involves relating the periodicities $p_\nu$ of the eigenspectrum to the number of exceptional points enclosed by the contour of real $\chi$, leading eventually to Eq.~(\ref{eq:znumber}).
In fact, based on the analysis by means of exceptional points, the number $\mathcal{Z}$ can be understood as an element of the $\mathbb{Z}_{2}$-group, where the nontrivial group operation is associated to the moving of an exceptional point across the real $\chi$ line.

Apart from indicating the mismatch between fractional charges and integer charge quantization, Eq.~\eqref{eq:znumber} emphasizes the importance of measuring at a sharp interface. As a reminder, in Sec.~\ref{subsec:braiding_in_FCS} we introduced the notion of spatially imprecise detectors, described by kernels of the form $e^{i\chi\widehat{n}p}W(\chi)e^{-i\chi\widehat{n}p}$, where measurements at the left and right interface occur with a probability of $p$ and $1-p$, respectively. Such kernels are no longer $2\pi$-periodic, and thus violate the assumptions to derive Eq.~\eqref{eq:znumber}. Indeed, while this unitary transformation leaves the eigenvalues invariant due to current conservation (as we already stated), it should become clear now that the geometric phases are affected by it, as they depend on the eigenvectors. In particular, for an arbitrary $p\in\mathbb{R}$, the kernel is not even periodic anymore, since the probabilities to measure from the left ($p$) and to the right contact ($1-p$) are incommensurable. Here, due to the incommensurability a meaningful geometric phase could only be defined by extending the system to two explicit counting fields, one on the left and one on the right, and considering generalizations of a 2D Chern number instead of the 1D geometric phase (in fact, very similarly in spirit to Ref.~\cite{Riwar_2016}). Such considerations of two-terminal measurements go beyond the scope here, and will be considered in the future. To sum up, we can only guarantee that $\mathcal{Z}$ is a meaningful topological number, when there is a local, sharp measurement at exactly one reservoir. This shows that $\mathcal{Z}$ is highly sensitive to how the transport is physically measured. Again we see, that the value of $\mathcal{Z}$ does a merely reflect the topology of the eigenspectrum $\lambda_\nu$, but is an additional number in its own right, putting the topology of $\lambda_\nu$ in relation with the global properties of $W(\chi)$. This dependence on whether or not a transport measurement occurs at a sharp interface will again become important in the subsequent Sec.~\ref{sec_analogy_fJE}, where we explain the analogy to the fractional Josephson effect.

We now relate the above geometric phase defined in the detector space to some of the other recent efforts to generalize geometric phases to open quantum systems. For instance,
the body of work elaborated in Refs.
\cite{Diehl_2011,Bardyn_2013,Budich_2015,Bardyn_2018} strives to
understand the topological structure of \textit{open} quantum systems
in terms of geometric phases related to the density matrix. In particular,
in Ref. \cite{Bardyn_2018} 1D open quantum systems with a momentum-like
degree of freedom $k$ are studied. The authors define a geometric
phase based on a many-body correlation function, which, in the thermodynamic limit, returns the standard Zak phase of the ground state in spite being at finite temperatures. A relation to the geometric phase defined in this paper can be found in so far as $\chi$ can be considered as a detector momentum, i.e., the conjugate variable to the number of transported electrons. 
However, the geometric phase
that we propose is a genuine nonequilibrium
quantity, and it is in general difficult to relate the connection $\sim\left(\nu\right|\partial_{\chi}\left|\nu\right)$
to the eigenvectors of a closed system and thus to the standard Zak
phase. 
A closer relation exists with respect to geometric
phases emerging in the FCS of time-dependently driven quantum pumps
\cite{Ning1992Oct,Landsberg1992,Sinitsyn_2007,Sinitsyn07PRL,Sinitsyn_2009,Ren2010Apr}. Further works by Refs. \cite{Pluecker_2016,Pluecker_2017}
elaborated that the pumping geometric phase is related to the gauge
degree of freedom arising from calibrating the transport detector.
The time-dependent driving manifests in an
explicit time-dependence of the kernel $W_{t}\left(\chi\right)$.
The resulting pumping geometric phase, $\sim\left(\nu\left(\chi\right)\right|\partial_{t}\left|\nu\left(\chi\right)\right)$,
provides the lowest order correction of the stationary contribution
to the FCS, due to nearly adiabatic driving. Importantly, the difference
to our approach is, that we perform a parallel vector transport not
along some pumping parameters, but along the counting field itself,
$\sim\left(n\left(\chi\right)\right|\partial_{\chi}\left|n\left(\chi\right)\right)$.
In essence, instead of a time-dependent driving, we here have to invoke the notion of a time-dependent measurement of the FCS. We will elaborate on this in more detail in Sec.~\ref{subsec:Measuring_geometric_phase}, where we examine strategies to measure $\mathcal{Z}$ experimentally.

\subsection{Analogy to fractional Josephson effect}\label{sec_analogy_fJE}

Here we compare the fractional effect in sequential electron tunneling to the fractional Josephson effect. Apart from the surprising fact that such a comparison is even possible in general, we have one more important motivation. Namely, as we mentioned in Sec.~\ref{subsec:periodicity_and_charge}, a proper requantization of charge seems highly nontrivial for Luttinger liquid theory. Therefore, an unambiguous comparison of the fractional nature of charges in the two systems, beyond the ad hoc arguments presented previously, seems difficult. However, as we will show now, in the case of superconducting transport across Josephson junctions, we can argue in a straightforward fashion, how charge quantization is preserved even in the topological regime. Thus, we can show that the same nontrivial quantized phase emerges, indicating a mismatch between the Cooper pair charge, the underlying unit of charge of the supercurrent, and the Majorana- and parafermions. This establishes, that the fractional nature of charge occuring in the fractional Josephson effect is the same as the one in sequential tunneling, except that the dynamics is completely quantum coherent, instead of dissipative.

To begin, let us note that in superconducting junctions, the number $\mathcal{N}$ of coherently transported Cooper pairs across the two superconductors has likewise its conjugate pendant in the form of the phase difference $\phi$. In fact, we can think of the pair $N$ and $\chi$ describing dissipative transport as the classical components of the quantum variables $\mathcal{N}$ and $\phi$. In particular, just as a $2\pi$-periodicity in $\chi$ indicates that the electron is the fundamental unit of dissipative electron transport, a $2\pi$-periodicity in $\phi$ corresponds to the Cooper pair being the underlying unit in which a supercurrent is mediated. However, excitations carrying a seeming fractional Cooper pair charge can appear in junctions with topological superconductors~\cite{Fu_2009,Zhang_2014,Orth_2015,Vinkler-Aviv_2017}. And once more, we need to define exactly, what is meant by fractional. The subgap transport physics of such junctions can be described by a Hamiltonian operator, $H_\text{sub}(\phi)$, valid for energies $<\Delta$, where $\Delta$ is the superconducting gap. Hence, very similarly to sequential electron tunneling, the transport can be described in terms of a linear operator along a transport degree of freedom ($\phi$). This Hamiltonian may likewise give rise to an eigenenergy spectrum with broken periodicity in $\phi$, either a $4\pi$-Josephson effect for Majorana fermions with half a Cooper pair charge~\cite{Fu_2009}, or an $8\pi$-Josephson effect for parafermion transport, carrying a quarter of the Cooper pair charge~\cite{Zhang_2014}. However, in the very same spirit of our above discussion in sequential tunneling, we stress that also here, care has to be taken as far as the periodicity of the Hamiltonian itself is concerned. We here advocate that the periodicity of the Hamiltonian be \textit{not} automatically given by the charge of the underlying quasiparticle or edge state, but (similarly to above) by the \textit{detector} - and that the choice of the periodicity of $H_\text{sub}$ (again, similar to sequential electron tunneling) has measurable consequences in the geometric phase defined along $\phi$. In particular, if an experimenter performs a subgap transport measurement by applying a bias across a sharp interface, he will measure the supercurrent response in units of Cooper pairs, requiring $H_\text{sub}$ to remain $2\pi$-periodic. That is, the periodicity of the underlying spectrum is broken \textit{with respect to} the Hamiltonian, indicating that the Cooper pair remains the underlying unit of the supercurrent, which will once more result in a nontrivially quantized geometric phase.

To visualize this at an example, let us consider the Hamiltonian put forth by Fu and Kane~\cite{Fu_2009}, describing a quantum spin Hall insulator (QSHI) proximitized by two superconductors, separated by a small non-proximitized patch, in the interval $0<x<L$,
\begin{equation}\label{eq_H_Fu_Kane}
\begin{split}
H=-iv_F\sigma_z\tau_z\partial_x-\mu\tau_z+M(x)\sigma_x\\+\Delta\theta(-x)\tau_x+\Delta\theta(x-L)e^{i\phi\tau_z}\tau_x\ ,
\end{split}
\end{equation}
where $\sigma_i$ and $\tau_i$ denote the left and right mover space of the QSHI edge states and the Nambu space, respectively. The term proportional to $M(x)$ has a finite support only within the interval $0<x<L$ and denotes a magnetic impurity which allows the spin-locked edge states to scatter. If we focus on the supercurrents (and neglect quasiparticle contributions), we can find a Hamiltonian describing the subgap physics, acquires a $2\pi$ periodic form from Eq.~\eqref{eq_H_Fu_Kane} (for a derivation using Beenakker's formula~\cite{Beenakker_1991}, see Sec.~\apptwotopJJ)
\begin{equation}\label{eq_H_eff}
H_\text{sub}(\phi)=\frac{\Delta}{2}\left(\begin{array}{cc}
0 & t_{e}\left[1+e^{i\phi}\right]\\
t_{e}^{*}\left[1+e^{-i\phi}\right] & 0
\end{array}\right)\ .
\end{equation}
For this Hamiltonian we find the eigenspectrum $\epsilon_\pm(\chi)=\pm \Delta |t_e|^2 \cos(\phi/2)$. Obviously, the eigenspectrum has a broken periodicity with respect to the Hamiltonian in Eq.~\eqref{eq_H_eff}, which is in fact $2\pi$-periodic by construction, as the phase is attached to only one of the superconducting contacts. However, could or should we have chosen a different representation for $H$, respectively $H_\text{sub}$, for instance with $\Delta e^{-i\phi/2}$ on the left and $\Delta e^{i\phi/2}$ on the right, or any other combination? In analogy to the unitary transformation of the kernel $W$, which expresses different measurement schemes (e.g., measuring with different probabilities $p$ and $1-p$ at the different contacts, see Secs.~\ref{subsec:braiding_in_FCS} and~\ref{sec_geometric_phase}), we find that also here, the answer depends on the measurement setup. Of course, as long as the total phase drop across the two superconductors is $\phi$, any choice would provide the same energy eigenspectrum. But different choices change the periodicity of the Hamiltonian. In particular, similarly to sequential electron tunneling, also here, the geometric phases picked up in a parallel vector transport across $\phi$ would differ for the different choices.

To fix the choice, we first consider how such a parallel vector transport is realized physically: namely, we can apply a voltage bias across the junction, such that $\phi\rightarrow\phi(t)=\phi_0+2eVt$. For sufficiently low voltages, the state of the quantum circuit simply evolves adiabatically along the same eigenvector, and picks up the geometric phase.
Consequently, the geometric phase depends on where the voltage drops within the sample. To convince ourselves that this conclusion does not break gauge invariance, note that there are in fact not two but \textit{three} terminals: in addtion to the two superconducting contacts, the QSHI itself has a chemical potential of its own, given by the term $-\mu\tau_z$. Hence, there are two independent chemical potential gradients, the one between the left superconductor and the QSHI and one between the QSHI and the right superconductor. Now we see that the representation in Eq.~\eqref{eq_H_Fu_Kane} is the correct choice, if we align the chemical potentials of both the left superconductor and the QSHI, such that $\mu=0$, while biasing the right superconductor with $V$. Here, Cooper pairs pick up the phase difference $\phi$ at one sharp interface (that is, sharp compared to $L$ and the coherence length), which is the interface between the QSHI and the right contact. We thus restored the Cooper pair as the fundamental unit of the supercurrent. The breaking of the $2\pi$-periodicity of the eigenspectrum expresses the fact that we need two consecutive tunneling events of Majorana fermions in order to transport one Cooper pair, in the same way that we need two consecutive sequential tunneling events in the single quantum dot, in order for the detector state to jump. And this charge mismatch is visible only, if the supercurrent is measured across a sharp interface. It is in this sense, that the fractionalization of charge is topologically of the same nature as in sequential tunneling. The subtleties related to comparing a nonequilibrium, nonhermitian system (described by $W$), with a Hermitian one (described by $H_\text{sub}$) will be discussed in a moment.

First however, note importantly, that the same argument holds for the $8\pi$ Josephson effect, which emerges when many-body interactions (in particular pair-backscattering) are present in the junction. The many-body interactions can be included on the level of Eq.~\eqref{eq_H_Fu_Kane} (of course in the appropriate many-body representation, that is, see e.g.,~\cite{Zhang_2014}), without changing the periodicity of the Hamiltonian. Thus we get a generally $2\pi$-periodic Hamiltonian with an $8\pi$-periodic spectrum, indicating that we need four consecutive parafermion tunneling events to register the transport of one Cooper pair. Note in particular, that for the $8\pi$ Josephson effect, Zhang and Kane present a subgap Hamiltonian with a periodicity of $8\pi$~\cite{Zhang_2014}. We on the other hand believe that the periodicity of the Hamiltonian should be chosen with care, taking into account the measurement setup - in particular, if the voltage drop occurs along a precise interface, the Hamiltonian should remain $2\pi$-periodic.

Crucially, once we have a description of the subgap physics through an operator $H_\text{sub}(\phi)$ which is detached from the continuum of states at energies $\geq\Delta$, the operator $H_\text{sub}(\phi)$ fulfils the exact same set of properties as $W(\chi)$, which are needed to compute the product of generalized geometric phases $\mathcal{Z}$ given in Eq.~\eqref{eq:znumber}, simply by replacing $\chi$ with $\phi$. Note that this renders our result of $\mathcal{Z}$ very general, as it can describe the topology of the subgap physics of quite generic Hamiltonians $H_\text{sub}$ (e.g., including many-body interactions), as long as the eigenspectrum is discrete, and detached from the continuum.

As indicated above, we have to address some important details related to comparing a nonequilibrium, non-Hermitian system, with a Hermitian system. Namely, while the kernel $W$ has in general complex eigenvalues, the spectrum of $H_\text{sub}$ is of course real, since supercurrents are an equilibrium property. Consequently, the connection of the latter to the braid group may not be obvious at first sight. As we show in Sec.~\apponeChermitianmatrices, the real spectrum with a crossing instead of a braid can be understood as a merging of two exceptional points with equal braid generator. Because the braid generators are equal, they cannot annihilate, and the nontrivial topology, including a nontrivial $\mathcal{Z}$ persists.
Therefore, we find the same $\mathcal{Z}$ for topological Josephson junctions, by replacing $\chi$ with $\phi$, and restricting ourselves to Hermitian operators. A nontrivial $\mathcal{Z}$ defined along $\phi$ indicates likewise the transport of a quasiparticle with a fractional Cooper pair charge, measured by a supercurrent mediated in units of integer Cooper pairs. In this sense, the product of geometric phases $\mathcal{Z}$ demonstrates a profound analogy between topological FCS in sequential electron tunneling and the fractional Josephson effect in topological superconducting junctions.

Note in addition that for sequential tunneling, no special symmtries are required to be in the topological phase. The only fundamental condition is a nonequilibrium bias, allowing for a complex, braided eigenspectrum, whereas in equilibrium the eigenspectrum of $W$ is real and no braiding can occur in general, as already pointed out by~\cite{Ren_2012} and in Sec.~\ref{subsec:explicit_examples}. Importantly, the same should also hold for the subgap physics in Josephson junctions, which are described through the Hermitian operator $H_\text{sub}$. The reason that the periodicity can nonetheless be broken on the level of the eigenspectrum of \textit{topological} superconducting junctions, is that additional symmetries are present which guarantee that the merging of braid generators remains stable. These symmetries are provided by the topological QSHI weak link, and the presence of magnetic impurities (for the $4\pi$ Josephson effect~\cite{Fu_2009}) or the presence of pair-backscattering in the weak link (for the $8\pi$ effect~\cite{Zhang_2014,Orth_2015}). As we see, the fractional effect in the Josephson junction relies on special symmetries, while the one in sequential electron tunnling is protected even in the absence of special symmetries, at the expense of rendering the dynamics non-Hermitian (thanks to the nonequilibrium bias).

In fact, this analogy opens up the possibility to simulate the topology of the fractional
Josephson-effect by means of regular electron transport. This idea
thus falls in line with a number of very recent proposals to implement
topological behavior known from the quantum domain through (semi)
classical dynamics, most notably the study of geometric and topological
effects in the diffusion dynamics of polymers \cite{Souslov_2017,Abbaszadeh_2017},
or also the implementation of the Su-Schrieffer-Heeger (SSH) model
through single electron transistors \cite{Engelhardt2017}. In addition, we emphasize that the here proposed simulator is potentially more stable than the 
fractional Josephson effect itself. In particular, since the Josephson effect relies on Cooper pair physics,
it is naturally susceptible to fermion parity breaking due to quasiparticle
tunneling \cite{Fu_2009,vanHeck_2011,Rainis_2012}. The incoherent fractional effect on the other hand, as discussed in Sec.~\ref{sec:braiding_and_frac_charge}, is \textit{defined}
in terms of single electron transport, and thus by nature insensitive to parity
breaking.

\section{Experimental verifications and possible applications\label{sec:Strategies-for-experiments}}

We have so far established the general concept of a fractional charge in sequential electron transport and its striking analogy to fractional charges in strongly correlated systems. While the charges of transport are defined through an analytic continuation of the eigenmodes of the moment generating function $m$, the global properties of $m$ are defined by the detector. We have then established that the mismatch of fractional charges with integer charge quantization lead to nontrivial geometric phases defined in the eigenspace spanned by the operator that generates the system-detector dynamics. Here we want to outline, how all these concepts can be verified experimentally, and how they could lead to possible applications.

First, we will briefly indicate that close to special symmetry points, a quasi-Poissonian transport of fractional charges can be measured by means of the lowest three cumulants in the long measurement time limit. Then, we argue that for a more general detection of fractional charges we need to extract time-dependent information of the detector, the waiting time distribution~\cite{Koch_2005,Brandes_2008,Welack_2009,Albert_2011,Rajabi_2013,Sothmann2014,Potanina_2017}. While it is known that the latter provides the eigenmodes, we here show that the waiting times provide also the geometric phases, and thus $\mathcal{Z}$. We will furthermore comment on the stability of the topological features with respect to detector errors and backaction. Finally, we will provide a concrete example how the braid phase transition in the FCS can be used to perform measurements beyond the detector resolution limit. 

\subsection{Experimental observation of ground state fractional charge}\label{sec_experiments_low_cumulants}

First, let us revisit the single-level quantum dot model, in order to propose concrete, feasible experiments to measure the fractional nature of the ground state charge, when the transport statistics is acquired  through a measurement of the current operator, that is, when measuring individual cumulants $C_k$. As we have already indicated in Sec.~\ref{subsec:explicit_examples}, the
topological phase is favoured when the transport is strongly biased
in one direction, e.g., from left to right, i.e., $\left(\mu_{\text{L}}-\epsilon\right)/k_{\text{B}}T\gg1$
and $\left(\epsilon-\mu_{\text{R}}\right)/k_{\text{B}}T\gg1$ (such
that $f_{L}\rightarrow1$ and $f_{R}\rightarrow0$).
Secondly, the presence of the fractional charge depends
also strongly on the coupling asymmetry. The most stable topological phase is reached for sufficiently symmetric coupling. In the regime of strong bias to the right, symmetric coupling corresponds to $\sigma\Gamma_\text{L}=\Gamma_\text{R}$. For strongly asymmetric coupling on the other hand, either $\sigma\Gamma_\text{L}\ll\Gamma_\text{R}$ or $\sigma\Gamma_\text{L}\gg\Gamma_\text{R}$, the spectrum is always trivial.

Let us here focus on a special regime of highly symmetric coupling. We will see now that the fractional charge is accessible through measuring only the lowest few cumulants. Namely, we find that the cumulant generating function is given as
\begin{equation}
c\left(\chi\right)\approx\gamma\left(e^{i\frac{\chi}{2}}-1\right)+\delta\gamma\left(e^{-i\frac{\chi}{2}}-1\right)\ ,\label{eq:quasi_poisson}
\end{equation}
where the term proportional to $\gamma=\left(\sigma\Gamma_\text{L}+\Gamma_\text{R}\right)/2$ is the leading term, and $\delta\gamma\ll\gamma_1$ is the lowest order correction.
The correction takes into account both small deviations from the exact tunnel coupling
symmetry, and small thermal excitations, $f_{\text{L}}\approx1-\delta f_{\text{L}}$
and $f_{\text{R}}\approx\delta f_{\text{R}}$. Taking the sum of these
two corrections, we find $\delta\gamma=\left(\sigma\Gamma_\text{L}-\Gamma_\text{R}\right)^{2}/\left(8\gamma\right)+\gamma\left(\sigma\delta f_\text{L}+\delta f_\text{R}/\sigma\right)/2$.
As we see, the statistics given in Eq.~(\ref{eq:quasi_poisson}) results in a nonequilibrium
transport statistics of ordinary electrons, which are indistinguishable
from an almost Poissonian transport statistics with a fractional
charge $e^{*}/e=1/2$. The leading term corresponds to the special limit which has already been remarked by~\cite{Nazarov_Blanter_2009}.
If the corrections were zero, it would suffice to measure the current $\left|I\right|=\frac{e}{2}\gamma$
as well as the noise $S=\left(\frac{e}{2}\right)^{2}\gamma$,
such that the Fano factor, as introduced in Eq.~(\ref{eq:Fano_factor}),
would provide us with a fractional charge $e^{*}/e=1/2$.

In general however, one cannot avoid having small deviations from
this sweet spot of exact symmetry and perfect bias, and
second term in Eq.~(\ref{eq:quasi_poisson}) cannot be neglected. Nonetheless, as we have argued at length in Sec.~\ref{subsec:explicit_examples}, the fractional charge is of topological nature, and thus protected against small perturbations. Therefore, these small deviations from the exact Poissonian limit do not destroy the fractional charge, but simply render the observation
of $e^{*}/e$ less straightforward. In order to detect it, we need the information of the higher cumulants. In particular, for the nearly Poissian limit, we
need in addition the third cumulant. From Eq.~(\ref{eq:quasi_poisson}),
we find that the relation between the cumulants is 
\begin{equation}
\frac{C_{k+1}}{eC_{k}}\approx\frac{e^{*}}{e}\left[1-2\frac{\delta\gamma}{\gamma}\left(-1\right)^{k}\right]\ ,\label{eq:near_poissonian_limit}
\end{equation}
up to first order in $\delta\gamma$. As we see, we have
two independent factors, the fractional charge $e^{*}/e$
and the correction factor $\delta\gamma/\gamma$.
To read them both out, we need to have access to the first,
second, and third cumulants, in order to create two independent ratios.
We stress the very different nature
of the two factors in Eq.~(\ref{eq:near_poissonian_limit}). While $e^{*}/e$ is given by the topology
and is thus precisely $=1/2$ for a large connected parameter subspace, the factor $\delta\gamma/\gamma$
depends on the device parameters. We therefore expect it to be possible
to test this result experimentally, e.g., through measuring the cumulants
for different parameter settings, and thus to demonstrate the topological origin
of $e^{*}/e$. We are confident that such experiments are within reach
\cite{Reulet_2003,Bomze_2005,Gustavsson_2006,Fujisawa_2006,Timofeev_2007,Sukhorukov_2007,Gershon_2008,LeMasne_2009, Flindt_2009,Ubbelohde_2012}. 

We note that in principle, the first term proportional to
$\gamma$ would contain a renormalization of the same order
as $\delta\gamma$. This correction is however irrelevant
when computing the respective ratios of the cumulants, as in Eq.~(\ref{eq:near_poissonian_limit}),
which is why we discarded it.

Finally, let us point out, that a similar regime can be reached for the double quantum dot level, see Fig.~\ref{figure_DQD}. For $f_\text{L}\approx 1$, $f_\text{R}\approx 0$, $f(\epsilon_\text{R}-\epsilon_\text{L})\approx 1$, and $\sigma\Gamma_\text{L}\approx\Gamma_{\text{LR}}\approx \Gamma_\text{R}$, we get a regime of Poissonian transport with a fractional charge $e/3$.


Beyond this long-time Poissonian regime, the measurement of $e_0^*$ would theoretically involve going to in principle arbitrary high cumulants, which seems unpractical. Moreover, for long measurement times, the decaying bands are suppressed. In the section that follows now, we will show how to measure ground state fractional charges beyond the Poissonian regime,
as well as the fractional charges associated to higher, decaying bands.

\subsection{Waiting time distribution and topology\label{subsec:waiting_time_distribution}}

In the existing literature, only indirect observations of topological
transitions in the FCS have been studied.
As we already stated earlier, in the limit of long measurement
times, $\tau\rightarrow\infty$, the moment generating function
for the long-time limit, as defined, e.g., in Eq.~(\ref{eq_m_disc}),
is globally discontinuous in the topological phase. The probing of nonanalytic
behaviour of the cumulant generating function has
been theoretically studied in Refs. \cite{Flindt_2013,Hickey_2014,Deger_2018} and
experimentally verified in Ref. \cite{Brandner_2016}. Here, the nonanalytic
behaviour in $c$ manifests as dynamical Lee-Yang zeros, and their
location can be extracted from a finite (short) time measurement of
high cumulants. Discontinuities in the cumulant generating function
do however not unequivocally probe a broken periodicity. First of
all, note that no actual braid phase transition is required for Lee-Yang zeros to
occur, see, e.g., Ref.~\cite{Flindt_2013} where very different models were considered
with a $c$ that is continuous with a discontinuous first derivative.
Within the specific context of sequential
electron transport, we can consider another simple generic example: imagine
two eigenmodes that do neither cross nor braid in the complex plane,
but where, within some connected region in $\chi$, the real parts
of the two eigenvalues change their order. At the points where the
crossing appears, $c$ is consequently discontinuous even in the absence
of a braid topology. Secondly, even if the nonanalytic behaviour indeed stems
from a nontrivial eigenspectrum, the mere presence of the discontinuity
does not allow us to extract the \textit{periodicity} of the eigenspectrum.
Nor does it provide information about the decaying modes which may
have a nontrivial charge of their own, as we have elaborated in the
previous sections. Therefore we conclude, that in order to measure
fractional charges, be it for the stationary or decaying modes, we
need the explicit information of the eigenmodes of the detector dynamics.

As was shown in Ref. \cite{Brandes_2008}, the eigenspectrum of $W\left(\chi\right)$
is accessible through the waiting time distribution of the detector,
when recording single tunneling events. For the sake of completeness,
we here briefly reiterate the main points. In a first step, to simplify the discussion, we limit our consideration to kernels that can be expressed as $W\left(\chi\right)=W_{0}+e^{i\chi}W_{+1}+e^{-i\chi}W_{-1}$. This form is sufficient to fully describe sequential electron tunneling.
Next, we suppose that the parts of the kernel which give rise to transport
can be written as products of vectors $W_{+1}=w_{+}\left|+\right)\left(+\right|$
and $W_{-1}=w_{-}\left|-\right)\left(-\right|$. The vector $\left|+\right)$
($\left|-\right)$) is the state the system is in
immediately after a ``click'' of the detector which increases (decreases)
the detector state by one. Note that these states do not have to be
pure states. The maps $\left(+\right|,\left(-\right|$ trace out
all the states the system can be in prior to the corresponding click, weighted with the appropriate probability. Note that in general $\left(+\right|\neq\left|+\right)^{T},\,\left(-\right|\neq\left|-\right)^{T}$. The \textit{experimentally measurable} quantities are then the the probabilities that time $\tau$ passes between the events $i$ and $j$, $(i|e^{W_0\tau}|j)$, where the indices can be $i,j\in\{0,+,-\}$. For instance $(\pm|e^{W_0\tau}|0)$ is the probability that time $\tau$ passes for the first click (either $+$ or $-$) to occur, when starting out in the stationary state. Similarly, $(+|e^{W_0\tau}|+)$ is the probability that time $\tau$ has elapsed between two $+$ clicks, without any other click occurring in between. Independently, one has to measure the rates $w_\pm$ at which such clicks occur. Beyond that, no further experimental input will be needed to compute the desired quantities, as we show in the remainder of this work.

Now we can formulate the time evolution of the system-detector dynamics as follows. We start from the full time evolution of the correlator
\begin{equation}\label{eq_Gij_time}
G_{ij}\left(\tau\right)=w_i\left(i\right|e^{W(\chi)\tau}\left|j\right)\ ,
\end{equation}
with $w_{0}=1$.
Expanding the time-evolution in orders of $w_\pm$,
we arrive at a Dyson-like equation of the form
\begin{equation}\label{eq:Dyson}
G_{ij}\left(\tau\right) =g_{ij}\left(\tau\right)+e^{i\chi}\sum_{k=\pm}\int_{0}^{\tau}dtg_{ik}\left(\tau-t\right)G_{kj}\left(t\right)\nonumber \ ,
\end{equation}
where the events that change the detector state correspond to the
vortices with the $e^{\pm i\chi}$ prefactors. The free correlator functions
$g_{ij}\left(t\right)=w_{i}\left(i\right|e^{W_{0}t}\left|j\right)$
are referred to as the waiting time distributions, again with $i,j\in\left\{ 0,+,-\right\}$.

Within this framework, the moment and cumulant generating functions are computed
through $m(\chi,\tau)=e^{c\left(\chi,\tau\right)\tau}=G_{00}\left(\tau\right)$. In
fact, in order to extract the eigenvalues $\lambda_{n}\left(\chi\right)$,
it suffices to perform a transformation into Laplace space, $g_{ij}\left(z\right)=\int_{0}^{\infty}d\tau e^{-z\tau}g_{ij}\left(\tau\right)$,
and solving Eq.~(\ref{eq:Dyson}) provides the condition
\begin{equation}
\det\left[\left(\begin{array}{cc}
g_{++}\left(z\right) & g_{+-}\left(z\right)\\
g_{-+}\left(z\right) & g_{--}\left(z\right)
\end{array}\right)^{-1}-\left(\begin{array}{cc}
e^{i\chi} & 0\\
0 & e^{-i\chi}
\end{array}\right)\right]=0\ .\label{eq:det_waiting}
\end{equation}
The set of $z$ which satisfies this condition corresponds to the
eigenspectrum $\left\{ \lambda_{n}\left(\chi\right)\right\} $. That
is, once the waiting time distribution, in form of the correlators $g_{ij}$, is measured and
analyzed in Laplace space, we have access to the full spectrum and
its topology, including the information of the charge. Of
course one remaining challenge is the proposition of feasible experimental
setups to actually measure the individual transport clicks. We will
present some ideas in Sec.~\ref{subsec:Practical-example}.

Finally, we note that to efficiently solve Eq.~(\ref{eq:det_waiting}),
we can use the fact that $W_{0}$ has an eigendecomposition of its
own, such that $g_{ij}\left(t\right)=\sum_{n}g_{ij,n}e^{l_{n}t}$,
where $\left\{ l_{n}\right\} $ is the eigenspectrum of $W_{0}$.
It then follows that $g_{ij}\left(z\right)=\sum_{n}g_{ij,n}\left(z-l_{n}\right)^{-1}$,
and consequently, solving Eq.~(\ref{eq:det_waiting}), eventually
simplifies to finding the zeros of a polynomial. On a practical side,
fitting the experimentally observed $g_{ij}\left(t\right)$ by a sum
of exponentials will necessarily come with errors. As long as these
errors are small, they cannot pose a serious threat to the correct
observation of the topology. We will elaborate on this with more detail
as soon as we propose explicit experimental examples, again, see
Sec.~\ref{subsec:periodicity_and_charge}.

For completeness, let us also mention a different method to measure the transport statistics of higher bands, proposed in Ref. \cite{Stegmann_2016b}. In essence, the
authors showed that the long-time cumulant generating function, $c(\chi,\infty)=\lambda_{0}\left(\chi\right)$,
can be used to infer the $\lambda_{n>0}\left(\chi\right)$ by means
of generalized factorial cumulants. This method could therefore potentially
be used to measure charges of the higher, decaying bands. However, we expect that we would yet again be limited to the nearly Poissonian regime of fractional transport. Moreover, we see no obvious path toward finding the geometric
phases of Eq.~(\ref{eq:znumber}). This is why we do not consider
it in detail in this paper. As we show now, the waiting times on the
other hand \textit{can} provide said geometric phases.

\subsection{Measuring geometric phases through waiting time distribution\label{subsec:Measuring_geometric_phase}}

Crucially, as indicated above, with the information provided by the waiting time distribution,
we can not only extract the \textit{eigenvalues} of $W\left(\chi\right)$,
but also the geometric properties of the \textit{eigenvectors}. In
order to appreciate this fact, it is important to realize that within
the framework of waiting times, Eq.~(\ref{eq:Dyson}), the counting
field $\chi$ enters simply as an external parameter, while the actual,
experimentally measured input is captured in the functions $g_{ij}\left(t\right)$.
That means, we can modify $\chi$ at will, including replacing it
with a time-dependent function of the form $\chi\rightarrow\omega t$.
We thus realize a time-dependent parallel vector transport along $\chi$-space.
Performing an adiabatic expansion of any $G_{ij}\left(\tau\right)$,
we find
\begin{equation}
G_{ij}\left(\tau\right)=\sum_{n}e^{\int_{0}^{\tau}dt\left[\lambda_{n,t}-\left(n_{t}\right|\partial_{t}\left|n_{t}\right)\right]}\left(i\left|n_{\tau}\right)\right.\left.\left(n_{0}\right|j\right)+\mathcal{O}\left[\frac{\left\Vert \omega\right\Vert }{\left\Vert \lambda\right\Vert }\right]\ ,\label{eq:G_adiabatic}
\end{equation}
where the generalized geometric phase $\int_{0}^{\tau}dt\left(n_{t}\right|\partial_{t}\left|n_{t}\right)$
appears, and the expansion parameter $\left\Vert \omega\right\Vert /\left\Vert \lambda\right\Vert $
is composed of the ratio between a driving rate $\left\Vert \omega\right\Vert \sim\left|\left(n_{t}\right|\partial_{t}\left|n_{t}'\right)\right|$
and a rate associated to the gap between eigenstates, $\left\Vert \lambda\right\Vert \sim\left|\lambda_{n,t}-\lambda_{n'.t}\right|$,
for $n'\neq n$. Equation (\ref{eq:G_adiabatic}) can be derived, starting from Eq.~\eqref{eq_Gij_time}, along similar lines as in Ref.~\cite{Mostafazadeh97}. A straightforward derivation can be done when discretizing time space; in fact, Eq.~\eqref{eq_geom_def} already provides the geometric part of Eq. (\ref{eq:G_adiabatic}). The additional dynamic prefactor $e^{\int_{0}^{\tau}dt\lambda_{n,t}}$ appears because the time evolution of $G_{ij}$ is generated by the full $W(\chi)$, see Eq.~\eqref{eq_Gij_time}. Equation (\ref{eq:G_adiabatic}) demonstrates that
we can access the information about geometric phases
by means of the correlators $g_{ij}$, importantly \textit{not} requiring
any additional effort on the experimental side apart from a measurement of the waiting times.

As we have pointed out in Sec.~\ref{sec_geometric_phase}, in order to extract the geometric phase from Eq.~(\ref{eq:G_adiabatic}), we need to close the path
in $\chi$. To repeat, for an eigenspectrum with broken $2\pi$-periodicity,
this means that for a given eigenvalue $n$, we set $\tau_{n}=\frac{2\pi}{\omega}p_{n}$,
such that we return to the original eigenvector, $\left|n_{\tau_{n}}\right)=\left|n_{0}\right)$,
and in this way, we receive the phase $e^{-\int_{0}^{\tau_{n}}dt\left(n_{t}\right|\partial_{t}\left|n_{t}\right)}=e^{-\int_{0}^{2\pi p_{n}}d\chi\left(n\left(\chi\right)\right|\partial_{\chi}\left|n\left(\chi\right)\right)}$.
Due to the already discussed redundancy, we change to the notation of the index of \textit{independent}
bands, $\nu$. The determination of the geometric phases and the topological number $\mathcal{Z}$,
Eq.~(\ref{eq:znumber}), is then straightforward. Since we have the entire information of the different $g_ij$ at disposal, we can compute the full eigenspectrum $\lambda_\nu(\chi)$ for all $\chi$. Based on this information, we can identify the individual contributions corresponding to a mode $n$ (respectively, $\nu$) in the sum of Eq.~\eqref{eq:G_adiabatic}. Finally, we can eliminate the overlap functions $\left(i\left|n_{\tau}\right)\right.\left.\left(n_{0}\right|j\right)$ by dividing $G_{ij}(2\pi p_\nu/\omega)$ with $G_{ij}(0)$. According to Eq.~\eqref{eq:znumber} the output from each mode $\nu$ is then multiplied. This procedure will be explicitly worked out at an example in the following section.

But before that, let us explain in more detail the notion of a time-dependent counting field. Time-independent counting fields
provide the \textit{time-averaged} current statistics. Time-dependent
counting fields, $\chi\rightarrow\chi\left(t\right)$, formally give
access to \textit{time-resolved} statistics \cite{Bednorz2008}.
To obtain information about a current measurement at a given time
$t$, one applies the functional derivative $\delta/\delta\chi\left(t\right)$
to the cumulant generating function. In this sense, in the standard
framework of time-resolved statistics, $\chi\left(t\right)$ is not
actually a mere function of $t$. Rather it is an assignment of a
\textit{different} counting field to each instance in time $t$.
In order to extract the geometric phase as proposed above, we implement
exactly this extra ingredient: with the above procedure, we essentially measure the transport at each time
$t$, and then assign a specific value for $\chi\left(t\right)$ ,
$\chi\rightarrow\omega t$. The function $G_{ij}$ can then be viewed
as a correlation of all measurements with this specific assignment.
Thus the here proposed procedure corresponds to a \textit{post-processing}
of the transport information we obtained through the measurement of
the waiting-time distribution, which is specifically designed to provide
us with the generalized geometric phase appearing in Eq.~(\ref{eq:G_adiabatic}).

Let us note a final crucial point. As we indicated, the derivation of Eq.~(\ref{eq:G_adiabatic})
is done in analogy to Ref.~\cite{Mostafazadeh97}, which
treats geometric phases in closed systems. However, there is an important difference
to the open system with a non-Hermitian matrix $W\left(\chi\right)$, which renders the extraction of the geometric
phase from above equation a bit harder. The issue is, that even though
$\omega$ can always be chosen sufficiently small such that the expansion
parameter $\left\Vert \omega\right\Vert /\left\Vert \lambda\right\Vert $
is $\ll1$ (provided that the spectrum is always gapped), nonadiabatic
correction can nonetheless become important, due to the prefactor $e^{\int_{0}^{\tau}dt\lambda_{n,t}}$,
or more specifically due to the eigenvalues $\lambda_{n}$ having
a non-positive real part. 

This fact becomes evident when considering the following. Suppose we start the
time evolution at time $t=0$ in the stationary state $\left|0\left(\chi=0\right)\right)$.
As time progresses, the dynamics stays mostly in the corresponding
eigenvector $\left|0\left(\omega t\right)\right)$, apart from some small
errors. If however the spectrum is nontrivial, there will come the moment,
where another eigenvector, say $\left|1\left(\omega t\right)\right)$,
will nominally become the new stationary state because $\text{Re}\lambda_{1}>\text{Re}\lambda_{0}$.
When this happens, the errors which are small in terms of the parameter
$\left\Vert \omega\right\Vert /\left\Vert \lambda\right\Vert $, will
receive a relative weight due to an exponential prefactor, here $e^{\int_{0}^{\tau}dt\left[\lambda_{1,t}-\lambda_{0,t}\right]}$,
which grows exponentially large over time. This makes the adiabatic
vector transport unstable. Moreover, if we perform the same expansion
right from the start in a decaying mode, say $\left|1\left(\chi=0\right)\right)$,
the same problem occurs even for a trivial spectrum, since all but
the stationary mode are being exponentially suppressed. Due to this
exponential growth, we should not think of Eq.~(\ref{eq:G_adiabatic})
as an actual approximation of $G_{ij}$ for small driving frequency
$\omega$, but rather as a formal expansion in the parameter $\left\Vert \omega\right\Vert /\left\Vert \lambda\right\Vert $,
without making any statement about whether the higher orders are negligible
or not.

Importantly, this does not hinder us from extracting the geometric
phases through Eq.~(\ref{eq:G_adiabatic}). The way out has implicitly already been mentioned before. Namely, as we stated already, through Eq.~(\ref{eq:det_waiting}), we already
have the full information of the individual $\lambda_{n}$ at each
$t$ at our disposal. As a consequence, when computing $G_{ij}$ for
a time-dependent counting field, $\chi=\omega t$, it is possible
to filter out at any time $t$ any specific band $n$, by identifying
it through its decay with $\lambda_{n}\left(t\right)$. Therefore, the same principle that is used to identify individual contributions of the single modes, serves at the same time to correct for unwanted transitions between bands. Once more,
we will show this is done in practice at the specific example which follows now. 

\subsection{Practical example of geometric phase measurement and robustness with
respect to errors and backaction\label{subsec:Practical-example}}

In the previous sections, we demonstrated the basic recipes of how
the topology in FCS can be experimentally accessed. In this
section, we elaborate on how these recipes can be implemented at a
specific practical example of a quantum point contact (QPC). We briefly show the extent to which the
observation of the fractional charges, and the corresponding geometric
phases, are stable with respect measurement errors and detector backaction.

As far as systematic errors are concerned, we have already commented on some aspects in Sec.~\ref{subsec:explicit_examples}. To repeat, in the quantum dot models we have dismissed levels and
charge states that are beyond the resonant energy window. Due to small temperatures,
these states are either mostly empty or mostly filled all the time, and therefore provide small corrections to
the current. These small corrections manifest themselves in the spectrum
of $W\left(\chi\right)$ as eigenvalues with only a very weak $\chi$-dependence.
They can therefore be considered as topologically inert eigenmodes, that do not
partake in any braid phase transition, and lead at most to small
corrections of the eigenspectrum. When accessing the detector dynamics through waiting times, such inert eigenvalues will appear as small corrections in the measured correlators $g_{ij}(\tau)$, and will likely remain undetected. To give an example, suppose that there are two important modes (just as in the single-level quantum dot), and a potentially arbitrary number of modes with only weak $\chi$-dependence. Then, $g_{ij}(\tau)=\sum_n g_{ij,n}e^{l_n \tau}$ will have two significant coefficients $g_{ij,0}$ and $g_{ij,1}$, and all other coefficients $|g_{ij,n\geq2}|\ll |g_{ij,0}|,|g_{ij,1}|$.

Similarly, detector backaction does not pose a significant thread for the topology. Detector-induced dephasing~\cite{Buks_1998} is obviously of no concern in the here studied purely classical dynamics. Apart from that, the QPC can induce inelastic transitions in the quantum system due to noise~\cite{Onac_2006,Gustavsson_2007}. However, as long as these processes are improbable (which can be assured by appropriately tuning the QPC parameters~\cite{Onac_2006}), they cannot pose any threat to the observation of the topological phases, as the latter are by definition stable with respect to small variations of the parameters. Furthermore, a finite reaction time of the QPC has been studied in~\cite{Flindt_2009}, which is however at least one order of magnitude faster than the tunneling dynamics. This additional effect would give rise to additional modes which are decaying very fast. Due to this separation of time scale, such modes can likewise not partake in a braid phase transition (see again our discussion in Sec.~\ref{subsec:explicit_examples}).

\begin{figure}
\includegraphics[width=1\columnwidth]{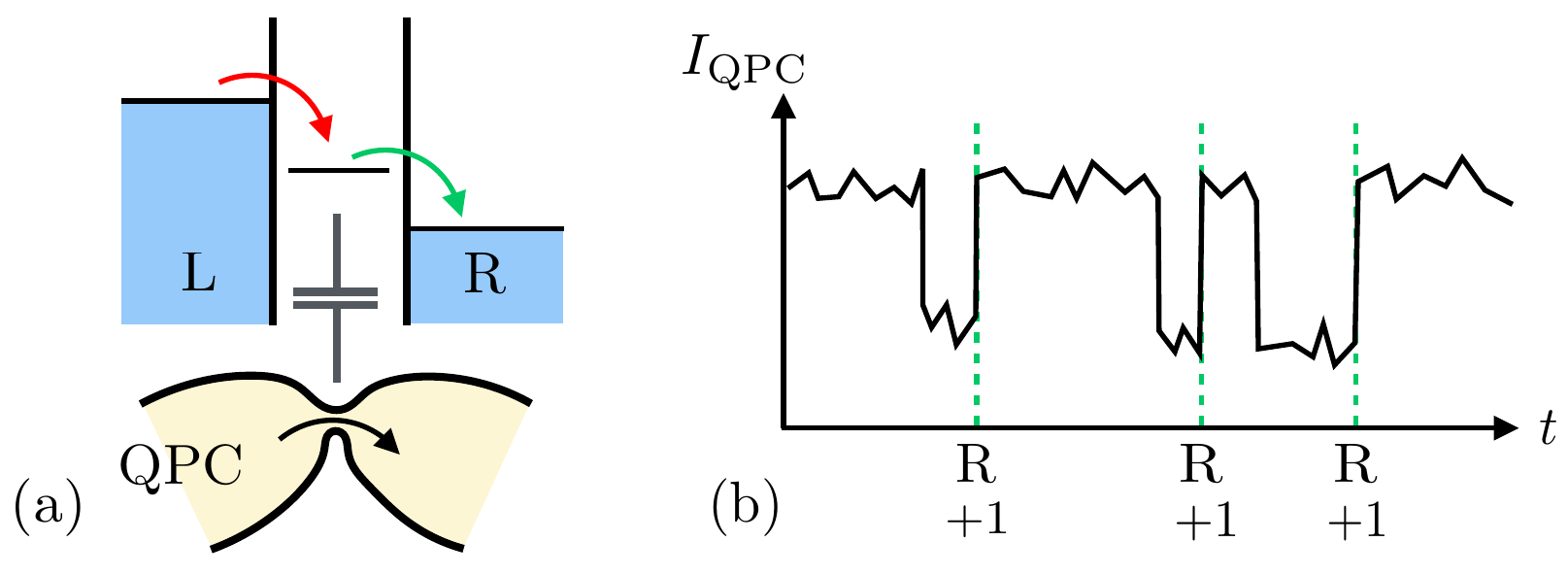}

\caption{(a) Measurement scheme to read-out the FCS of a single-level quantum dot. A
quantum point contact (QPC) is capacitively coupled to the dot in order
to time-dependently resolve the charge state. We assume a setup in
which the chemical potentials are biased such that an electron entering
the dot comes most likely from the left reservoir, whereas the electron
leaves most likely to the right. (b) The current in the QPC as a function
of time. Depending on the charge state on the quantum dot, the current is either
reduced (filled dot) or not (empty). Since we are interested in the
transport statistics into the right lead, we discard the charging
events, and only record the decharging events. That is, each time the QPC current rises, we increase the detector state by one (indicated with the
green, dashed line).\label{fig:QPC_FCS_measurement}}
\end{figure}

Apart from that, there can also occur small errors due to the specific measurement setup, as we will discuss now.
For the sake of concreteness, we consider again
the simple single-level quantum dot model from Sec.~\ref{subsec:explicit_examples},
adding an explicit detector scheme. Since we believe that an actual
measurement of electrons leaving or entering the right lead could
be experimentally challenging, we propose to use a highly simplified measurement
setup with a quantum point contact (QPC) capacitively coupled to the
quantum dot, see Fig.~\ref{fig:QPC_FCS_measurement}a, as it has been deployed in Ref.~\cite{Flindt_2009}. The current
flowing through the QPC is sensitive to the charge state of the quantum
dot, see Fig.~\ref{fig:QPC_FCS_measurement}b. In the chosen example,
the QPC has per se no direct means to distinguish the direction in
which an electron tunnels. However, we can consider a regime where the chemical potentials are biased such that the charging and
decharging events can with very high probability be associated to
a charge transfer from the left, and to the right, respectively. In
order to evaluate the transport statistics to the right, we can thus
discard the charging events monitored by the QPC and only record the
decharging events (as shown in Fig.~\ref{fig:QPC_FCS_measurement}b).
In this way, the detector will miss the very unlikely events of thermal
excitation of electrons against the bias, which thus represents a
first (small) source of measurement errors.

Consequently, the kernel defined in Eq.~(\ref{eq:kernel_single})
has to be replaced by the new kernel
\begin{equation}
W_{\text{QPC}}\left(\chi\right)=\sum_{\alpha}\left(\begin{array}{cc}
-2\Gamma_{\alpha}f_{\alpha} & \Gamma_{\alpha}\left[1-f_{\alpha}\right]e^{i\chi}\\
2\Gamma_{\alpha}f_{\alpha} & -\Gamma_{\alpha}\left[1-f_{\alpha}\right]
\end{array}\right)\ ,\label{eq:W_QPC}
\end{equation}
that is, we measure the topology of $W_{\text{QPC}}$ rather than
$W$ from Eq.~(\ref{eq:kernel_single}). For the extreme limit $f_{\text{L}}\rightarrow1$
and $f_{\text{R}}\rightarrow0$, $W_{\text{QPC}}\left(\chi\right)$
reduces to Eq.~(\ref{eq:kernel_single}), thus, $W_{\text{QPC}}\left(\chi\right)$
and $W\left(\chi\right)$ are equivalent. For finite thermal processes,
the two differ slightly.

Along the lines of Sec.~\ref{subsec:waiting_time_distribution}, we
can extract the eigenvalues $\lambda_{n}\left(\chi\right)$ of $W_{\text{QPC}}\left(\chi\right)$
and the topology of the eigenspectrum through measuring the waiting
time distribution between two decharging events, $g_{++}\left(t\right)$.
Since we have only the addition of charges to the right lead as a
measured transport event, Eq.~(\ref{eq:det_waiting}) simplifies to
$g_{++}^{-1}\left(z\right)-e^{i\chi}=0$, to determine $\lambda_{n}\left(\chi\right)$. 

In addition, we may determine the geometric phase due to a parallel
vector transport, as introduced previously in Sec.~\ref{subsec:Measuring_geometric_phase}.
For this simple example, we now present the strategy to filter out
a specific band, in order to avoid the adiabatic instability problem
mentioned in Sec.~\ref{subsec:Measuring_geometric_phase}. To this
end, we take the entire time interval $\tau$, and discretise it,
$\tau=\Delta t_{0}+\Delta t_{1}+\ldots+\Delta t_{M}$. For each time
interval $\Delta t_{m}$, where $m$ is integer and $0\leq m\leq M$,
we assign a constant counting field $\chi_{m}$. This enables us to
solve Eq.~(\ref{eq:Dyson}) piecewise exactly. We demonstrate the
full derivation in Sec.~\appthreegeometricfromwaiting.
Here we merely recapitulate the main steps. Note that we focus on $G_{++}$, since it is the only relevant correlator for the here considered detection scheme.

As it turns out, it is particularly useful to transform to Laplace
space for each time interval. We define the operator for this transformation as $\mathcal{L}_{M}=\prod_{m=0}^{M}\int_{0}^{\infty}d\Delta t_{m}e^{-z\Delta t_{m}}$.
This enables us to find a closed form for $G_{++}\left(z_{0},z_{1},\ldots\right)=\mathcal{L}_{M}G_{++}\left(\Delta t_{0},\Delta t_{1},\ldots\right)$.
Now, the crucial step follows. We can compute the geometric phase as $B_\nu=g_{++}^{-1}[\lambda_\nu(0)]\prod_{m=0}^{M}\left[z_{m}-\lambda_{\nu}\left(\chi_{m}\right)\right]G_{++}\left(z_{0},z_{1},\ldots\right)$
and subsequently take the limit of $z_{m}\rightarrow\lambda_{\nu}\left(\chi_{m}\right)$.
In this way we filter out the contribution of the band of interest,
$\nu$, avoiding the instability occuring in the adiabatic time evolution.
In the present example, we find that for $\nu=0$ the geometric phase
$\mathcal{B}_{0}=e^{-\int_{0}^{2\pi p_{0}}d\chi\left(0\right|\partial_{\chi}\left|0\right)}$
is given as (see Sec.~\appthreegeometricfromwaitingshort)
\begin{align}
\mathcal{B}_{0} & =g_{++}^{-1}\left[\lambda_{0}\left(0\right)\right]\lim_{M\rightarrow\infty}\prod_{m=0}^{M-1}g_{++}\left[\lambda_{0}\left(\chi_{m}\right),\lambda_{0}\left(\chi_{m+1}\right)\right]\nonumber \\
 & \times e^{i\chi_{m}}\frac{\left[\lambda_{0}\left(\chi_{m}\right)-l_{0}\right]\left[\lambda_{0}\left(\chi_{m}\right)-l_{1}\right]}{\lambda_{0}\left(\chi_{m}\right)-\lambda_{0}\left(\chi_{m}+2\pi\right)}\ ,\label{eq:B_0}
\end{align}
with $\chi_{m}=2\pi p_{0}m/M$, and $g_{++}\left(z,z'\right)=\sum_{n=0,1}g_{++,n}\left(z-l_{n}\right)^{-1}\left(z'-l_{n}\right)^{-1}$.
Importantly, Eq.~(\ref{eq:B_0}) explicitly shows that $\mathcal{B}_{0}$
can be expressed fully in terms of the function $g_{++}\left(z\right)$
and the eigenvalues $l_{0,1}$ of $W_{0}$, and is consequently directly
accessible through the Laplace analysis of the waiting time distribution. 

\begin{figure}
\includegraphics[width=0.7\columnwidth]{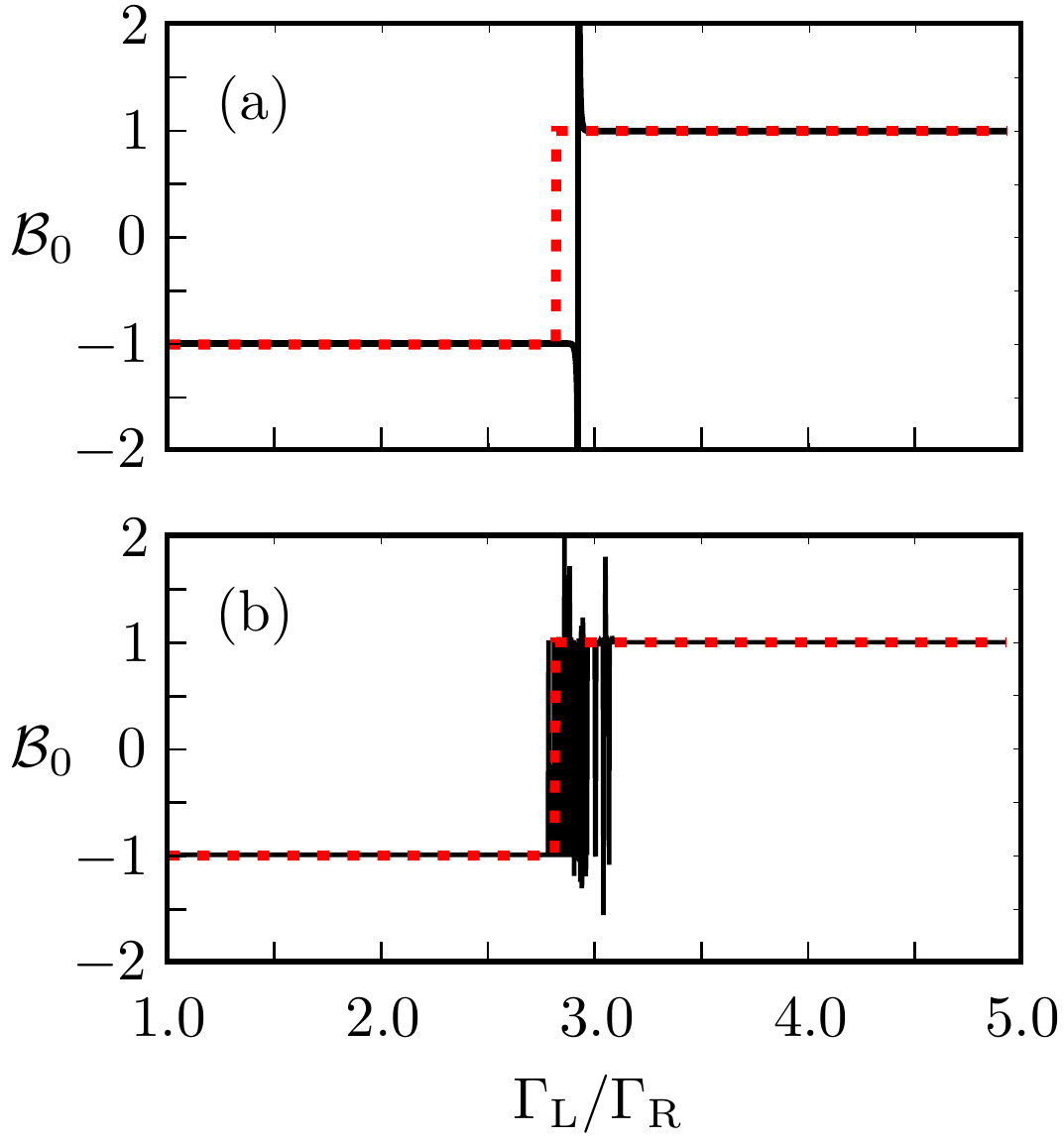}

\caption{The geometric phase $\mathcal{B}_{0}$ as a function of $\Gamma_{\text{L}}/\Gamma_{\text{R}}$
for some error sources. The red dashed line corresponds to the
exact $\mathcal{Z}$-number that would be expected for the quantum
dot model if the measurement were ideal. In (a) the black curve corresponds to the case where an
error occurs due the inaccurate measurement setup (see Fig.~\ref{fig:QPC_FCS_measurement}).
Near the transition $\mathcal{B}_{0}$
becomes ill-defined. This is due
to the numerical evaluation with finite elements, $\Delta\chi$. For
$\Delta\chi$ approaching $0$, the divergence dissappears, see also main text. In (b)
we show (again in black) the curve due to fitting errors when extracting
the parameters $g_{++,0},g_{++,1}$ and $l_{0},l_{1}$ from the waiting time distribution.
In general, we observe that while errors give rise to faulty behaviour
near the transition, the results are extremely stable away from the
transition, thus allowing for a reliable measurement of topological
numbers through waiting times. The parameters in both figures are
$\mu_{\text{L}}-\epsilon=5k_{\text{B}}T$, $\mu_{\text{R}}-\epsilon=-4k_{\text{B}}T$.
\label{fig:geometric_phase_waiting_times}}

\end{figure}

In Fig.~\ref{fig:geometric_phase_waiting_times}, we show the result
of $\mathcal{B}_{0}$, as a function of $\Gamma_{\text{L}}$. That
is, we assume that the experimenter can modify the tunnel coupling
to at least one of the leads, and for each constellation of $\Gamma_{\text{L}}/\Gamma_{\text{R}}$,
evaluates the waiting time distribution of the detector.  The red
dashed line shows $\mathcal{B}_{0}$ for the original model given
in Eq.~(\ref{eq:kernel_single}), i.e., assuming ideal measurement.
In Fig.~\ref{fig:geometric_phase_waiting_times}a, the black line
corresponds to $\mathcal{B}_{0}$ for the simplified measurement scheme
with the QPC, described in above Eq.~(\ref{eq:W_QPC}), and evaluated
through the waiting time distribution, according to Eq.~(\ref{eq:B_0}).
Firstly, we observe that $\mathcal{B}_{0}$ evaluated through waiting
times returns either $\pm1$. The fact that already $\mathcal{B}_0$ alone is quantized, and we do not need to take the product $\mathcal{Z}=\prod_\nu\mathcal{B}_\nu$, is due to the very simplified system with only two available states, and a detector only clicking in one direction ($+$). For a more complicated system, we would have to consider the product $\mathcal{Z}$ to obtain a quantized number. Note that very
close to the transition, we find that $\mathcal{B}_{0}$ does not return $\pm1$, and instead becomes ill-defined. This is due to the fact, that the evaluation
was done in a discretized way, with a finite resolution in $\chi$,
where $M$ is finite. In the limit $M\rightarrow\infty$, the function
approaches the exact step function. Secondly, we see that the transition
from topological, $\mathcal{B}_{0}=-1$, to trivial, $\mathcal{B}_{0}=+1$,
is shifted and occurs at a slightly different ratio of $\Gamma_{\text{L}}/\Gamma_{\text{R}}$.
This is because the setup with the QPC is missing the thermally excited
processes. Thus the detector overestimates the current, which renders the topological phase more stable (i.e., the system is in the topological phase for a larger parameter subspace). We can similarly expect that for another type of error, which underestimates the current, the effect would be the opposite, that is, a less stable topological phase.

Finally, we comment on errors in the fitting process. Namely,
we may imagine that for each setting of $\Gamma_{\text{L}}/\Gamma_{\text{R}}$,
the experimenter performs a new evaluation of the detector dynamics,
and fits the function $g_{++}\left(\tau\right)$ with a double exponential
to extract the indices $g_{++,0},g_{++,1}$ and $l_{0},l_{1}$. It
is then reasonable to assume that the fitting process might be subject
to fluctuations for each setting. We here model these fluctuations
as follows. We supplement the four extracted parameters, $g_{++,n}$
and $l_{n}$, each with an independent small error, $g_{++,n}+\delta g_{++,n}$
and $l_{n}+\delta l_{n}$, which we update with pseudo random numbers
for each value of $\Gamma_{\text{L}}/\Gamma_{\text{R}}$. In Fig.
\ref{fig:geometric_phase_waiting_times}b, we choose the error magnitude
such that both $\delta g_{++}$ and $\delta l$ change within $10\%$
of the original value $g_{++}$ or $l$, respectively. We see that
close to values of $\Gamma_{\text{L}}/\Gamma_{\text{R}}$ where the
transition occurs in the ideal case, the fluctuations lead to a random
back and forth switching between $\mathcal{B}_{0}=+1$ and $-1$.
This is because in this critical region, a small change in the $g_{++}$'s
and $l$'s can undo, respectively, redo the braid phase transition.
Far away from this region, the topological phases remain stable.

Overall, we have examined backaction as well as some realistic errors due to the concrete measurement setup, none of which can destroy the braid phase transition
in the eigenspectrum, nor hinder the observation of the associated
geometric phase $\mathcal{B}$. As we showed at some examples, errors merely meddle with the precise
measurement of the location where the topological transition appears.
We further note, that in above examples, we went to rather high error
amplitudes, in order to render the effects visible in Fig.~\ref{fig:geometric_phase_waiting_times}.
In reality (see also Ref.~\cite{Flindt_2009}), we are confident that those errors can be reduced significantly.

In fact, let us note at this point, that in Ref.~\cite{Flindt_2009}, the waiting time distribution of a quantum dot coupled to a QPC has indeed been experimentally measured. Their setup thus would allow for the extraction of this quantized phase. Unfortunately, the raw data of the waiting time distribution is not available from Ref.~\cite{Flindt_2009}, which is why we cannot directly extract the geometric phases $\mathcal{B}_n$ and their product $\mathcal{Z}$. However, considering the values of the tunneling rates they extract, they find the ratio $\Gamma_\text{L}/\Gamma_\text{R}=0.25$. Thus, their system should actually be in the topological phase with $e_0^*=e/2$ and $\mathcal{Z}=-1$. 
In order to observe a transition from topological to trivial, the waiting times would have to be extracted for different ratios of $\Gamma_\text{L}$ and $\Gamma_\text{R}$, which was not done in Ref.~\cite{Flindt_2009}.

\subsection{Going beyond the resolution limit\label{subsec:beyond_resolution_limit}}

\begin{figure}
\includegraphics[width=1\columnwidth]{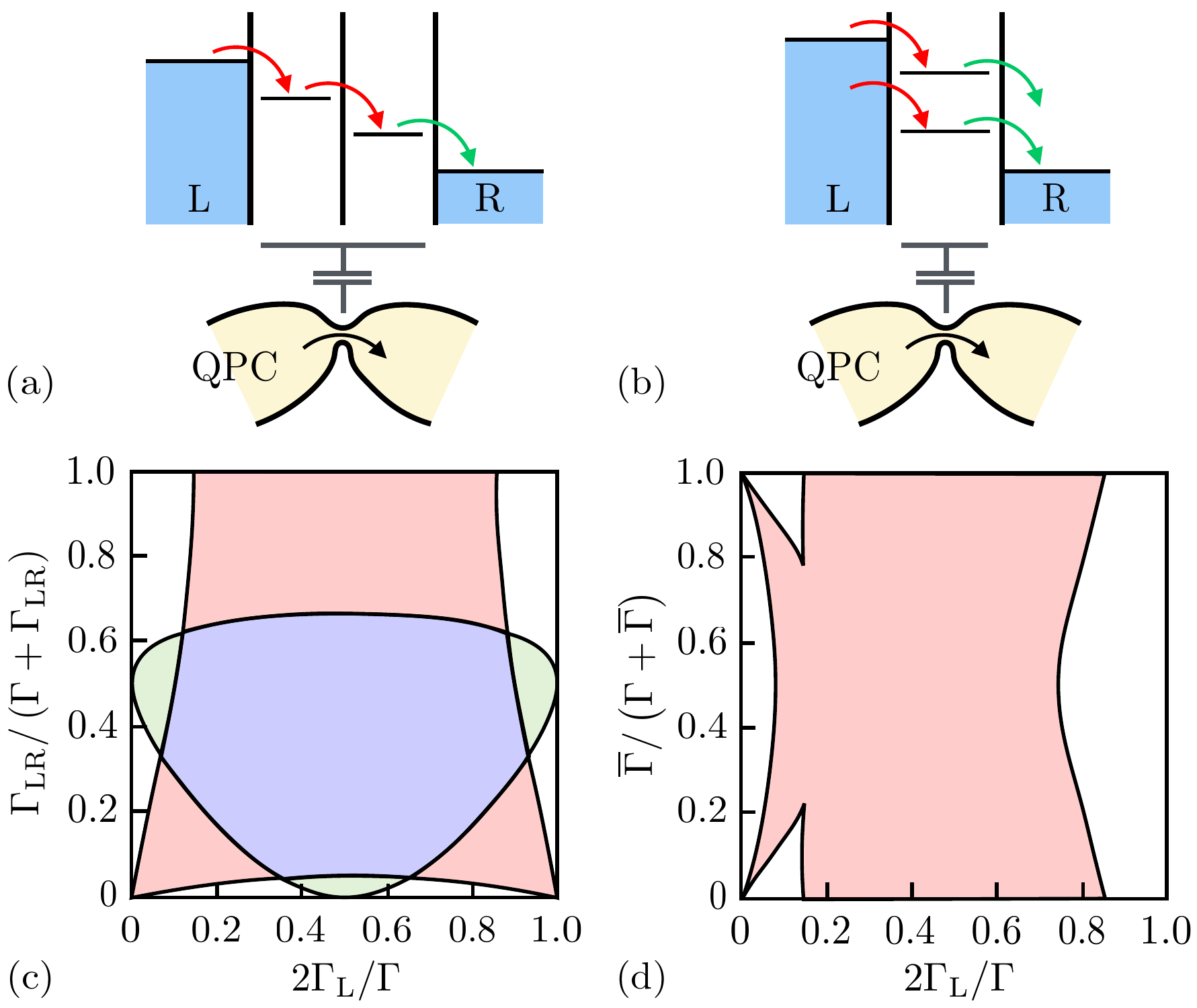}

\caption{Transport models to illustrate the principle of resolving processes beyond the detectors resolution limit. The electrons may tunnel through two levels in either a serial (a) or
a parallel (b) configuration. The statistics are measured through
a QPC which cannot resolve individual charge islands. In (c) and (d)
we show the topological phase diagram for the serial and parallel
configuration, respectively, as a function of the tunnel coupling
parameters. We defined $\Gamma=2\Gamma_{\text{L}}+\Gamma_{\text{R}}$
and $\overline{\Gamma}=2\overline{\Gamma}_{\text{L}}+\overline{\Gamma}_{\text{R}}$.
On the x-axis of both figures, we change the left and right asymmetry,
$2\Gamma_{\text{L}}/\Gamma$, where for (d) we assume for simplicity
$2\Gamma_{\text{L}}/\Gamma=2\overline{\Gamma}_{\text{L}}/\overline{\Gamma}$.
On the y-axis of (c) we modify the relative tunneling rate of the
inelastic process, $\Gamma_{\text{in}}/\left(\Gamma+\Gamma_{\text{in}}\right)$,
whereas in (d) we modify the relative total tunneling rate for each
level, $\overline{\Gamma}/\left(\Gamma+\overline{\Gamma}\right)$.
In (c) and (d) the white areas correspond to a fully trivial eigenspectrum
of $W\left(\chi\right)$. The red area corresponds to the same topological
phase as in Fig.~\ref{figure_DQD}c, with a braided stationary mode
with a fractional charge $e_0^{*}=e/2$. The blue and
green areas indicate the topological phases as shown in Fig.~\ref{figure_DQD}b
and Fig.~\ref{figure_DQD}d, respectively, and occur only in the serial
level configuration. In particular the blue phase, where all three eigenvalues merge to a band with charge $e_0^{*}=e/3$, is stable
for a large parameter space, and provides a unique signature of the
presence of two serially coupled islands instead of one. \label{fig:QPC_beyond_resolution}}

\end{figure}

As we have argued in Sec.~\ref{subsec:periodicity_and_charge}, we can directly link the mismatch between fractional charges and integer charge quantization, and the resulting periodicity breaking due to a braid phase
transition, to a detector measuring a stochastic process below its
resolution limit. We expect that the latter provides new applications for measuring techniques. To demonstrate this principle at work, we consider an extension of the previously introduced
simple QPC setup (see Fig.~\ref{fig:QPC_FCS_measurement}). Namely, we can add a second
level to the system, where for simplicity, we consider two extreme
cases. We either consider again the serial double quantum dot model
from Fig.~\ref{figure_DQD} with a QPC, which is now coupled equally
to both islands, see Fig.~\ref{fig:QPC_beyond_resolution}a. This
corresponds to an archetype example of a detector that cannot resolve
the two individual islands.

We put this system into relation with another one, containing only one charge island (a SQD), however with \textit{two} available levels instead of one. Importantly, note that if we were to compare the serial DQD to a SQD with only one available level, as in Fig.~\ref{figure_SQD}, we would not need any sophisticated topological argument to differentiate the two systems, as we could already infer the presence of the second dot simply by counting the number of degrees of freedom in the waiting time distribution: in the SQD, we have only two dominant eigenvalues, whereas in the DQD we have three. The distinction between a single and a double island however is less straightforward, if the single island has two levels, which are not coupled in series, but
in parallel, see Fig.~\ref{fig:QPC_beyond_resolution}b. Here, both systems have three dominant eigenmodes. We tune yet
again the energy levels and charging energies, such that only one
extra electron can enter the quantum system, and electron transport
goes with very high probability towards the right. Once more, we neglect
any spurious thermal errors, i.e., thermally excited processes
going against the energy gradient (as the topology is stable with respect to such small errors). As a consequence, we can write
the kernel of the serial model simply as
\begin{equation}
W_{\text{s}}\left(\chi\right)=\left(\begin{array}{ccc}
-2\Gamma_{\text{L}} & 0 & \Gamma_{\text{R}}e^{i\chi}\\
2\Gamma_{\text{L}} & -\Gamma_{\text{LR}} & 0\\
0 & \Gamma_{\text{LR}} & -\Gamma_{\text{R}}
\end{array}\right)\ ,
\end{equation}
where the factor $e^{i\chi}$ takes again into account that we only record the decharging
events in $I_{\text{QPC}}$. As for the parallel configuration, we find
\begin{equation}
W_{\text{p}}\left(\chi\right)=\left(\begin{array}{ccc}
-2\Gamma_{\text{L}}-2\overline{\Gamma}_{\text{L}} & \Gamma_{\text{R}}e^{i\chi} & \overline{\Gamma}_{\text{R}}e^{i\chi}\\
2\Gamma_{\text{L}} & -\Gamma_{\text{R}} & 0\\
2\overline{\Gamma}_{\text{L}} & 0 & -\overline{\Gamma}_{\text{R}}
\end{array}\right)\ ,
\end{equation}
where the third row (column) account for the coupling with the second
level at higher energies, which are coupled through the rates $\overline{\Gamma}_{\text{L}},\overline{\Gamma}_{\text{R}}$.

Crucially, while the two systems cannot be distinguished by the number of degrees of freedom, they can be distinguished based on the topology of the QPC dynamics.
In fact, the topological phases that occur in the serial configuration
were analyzed already in Sec.~\ref{subsec:explicit_examples}. In
Fig.~\ref{fig:QPC_beyond_resolution}c, we show the phase diagram
as a function of the tunnel coupling rates. In particular, there is
a very stable phase with three braided eigenvalues (blue area) for a large parameter space, with a spectrum of
periodicity $p_0=3$ and thus the charge $e/3$, see also Fig.~\ref{figure_DQD}b. Through making the tunneling rates more
and more asymmetric, the system-detector dynamics changes to the simpler braid topology
with a spectrum as in Figs.~\ref{figure_DQD}c and d (red and green areas, respectively), and finally
to a trivial phase (white area).

\begin{figure}
\includegraphics[width=1\columnwidth]{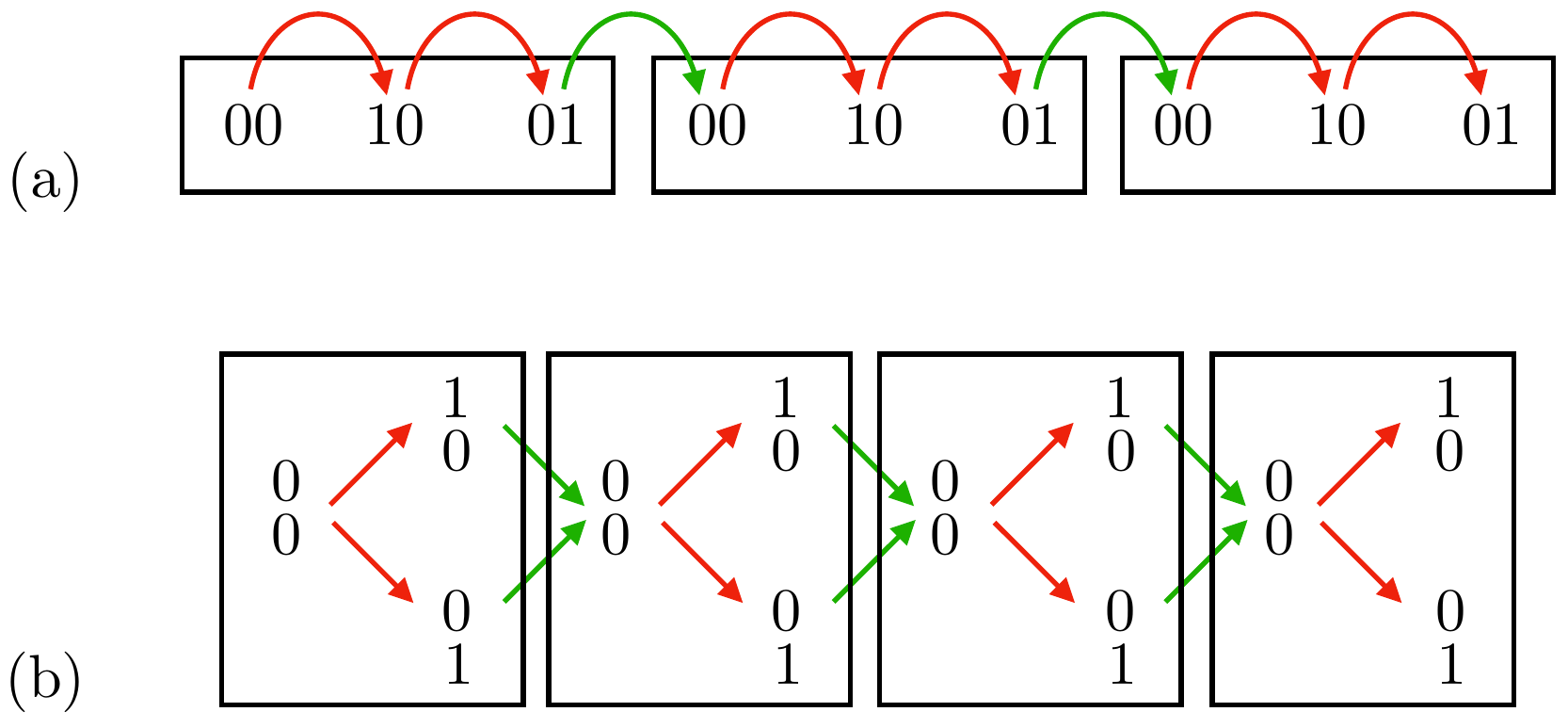}
\caption{Representation of the system-detector dynamics of the transport through a serial (a) and a parallel (b) level configuration. The black rectangles represent the detector pixels. Only processes that change between different pixels are the ones registered by the QPC (green arrows), the others are undetected (red arrows). In the serial configuration (a) in each pixel, there are three available states, either both quantum dots empty ($00$), or one electron in the left ($10$) or the right dot ($01$). In order to move to the next pixel, we need to perform three consecutive tunneling events. While in the parallel configuration (b) there are likewise three states within one pixel (either both levels empty, or either the lower or the upper level filled with one electron), one needs only two consecutive processes to change the detector pixel. Therefore, the $e/3$ phase is impossible in the latter case. }
\label{fig_beyond_resolution_explanation}
\end{figure}

Importantly, the $e/3$ phase occurs exclusively in the serial configuration.
In the parallel configuration, see Fig.~\ref{fig:QPC_beyond_resolution}d,
the QPC dynamics are either trivial (white area) or have a braided
stationary mode with charge $e_{0}^{*}=e/2$ (red area).
No other topological phases can be observed in spite of the presence
of a second available level. The reason for the absence of the $e/3$ phase can be explained intuitively, when representing the processes in a graph including the detector states $N$, see Fig.~\ref{fig_beyond_resolution_explanation}. While the serial DQD can be represented as a process with a step size three times smaller than the detector resolution (Fig.~\ref{fig_beyond_resolution_explanation}a), the same cannot be accomplished in the parallel configuration. For the latter, there are only two consecutive processes needed to move to the next detector pixel (Fig.~\ref{fig_beyond_resolution_explanation}b). We conclude that the occurrence of the
$e/3$ phase is a direct consequence of the fact that in the serial
configuration an electron needs to execute three consecutive tunneling
events. Thus we find a unique topological signature in the statistics
of a QPC detector, which enables us to determine the presence or absence of a second charge island for a large parameter regime, 
even if the detector cannot cannot distinguish the islands. As we have already shown
in the previous section, small errors and deviations from the ideal
models considered here, cannot destroy the effect.

Finally, we anticipate that the braid topology could potentially
be useful well beyond the context of quantum transport, as a general
tool to observe the dynamics of very small, discrete processes beyond
the detectors resolution limit. In above example, and in Sec.~\ref{subsec:periodicity_and_charge},
we have so far focussed on the limit of a detector resolution which
is commensurable with the step size of the discrete process. In the
general discussion in Sec.~\ref{subsec:periodicity_and_charge}, this
fact was expressed in the detector resolution $\mathcal{A}$ being
an integer multiple of $e^{*}$. In the concrete example here, we
have considered one QPC which measures two dots (or charge islands).
In order to generalize this idea, we believe it is of very high interest
to include the possibility of a detector resolution $\mathcal{A}$
that is incommensurable with the step size of the observed discrete
process. This question goes however well beyond the scope of the present
work, and will be pursued in the future.

\section{Conclusion}

In this work, we have shown that a topological transition in the
system-detector dynamics leads to FCS with eigenmodes carrying fractional charges. This realization relies on the fact that the fundamental integer charge quantization fixes the global properties of the moment generating function, and on the proposition that fractional charges can only be well-defined by means of an analytic continuation of the eigenmodes of the moment generating function. In particular, we showed that we can map the dynamics of topological FCS in a very generic transport situation to a hypothetical system with fractionally charged quasiparticles, supplemented with a charge detector with integer charge resolution, ensuring the indivisibility of the elementary charge. Our result therefore raises the question how far the transport signatures of fractionally charged quasiparticles are unique to exotic excitations, such as Luttinger liquid or Laughlin quasiparticles.

We have further demonstrated
the existence of a generalized geometric phase, which gives rise to
a topological number directly indicating the mismatch between fractional charges and integer charge quantization. Moreover, we have shown that this topological number allows
us to establish a profound analogy to the topology of the fractional Josephson
effect in superconducting junctions, thus comparing a system with fully dissipative transport to a system with coherent supercurrents. This strenghtens our proposed indistinguishability of fractional charges in sequential electron tunneling and fractionally charged quasiparticles, and provides in addition an unexpected possibility to simulate crucial aspects of the transport physics of strongly correlated quantum systems by means
of classical, incoherent dynamics.

We have provided explicit strategies to verify the various claims in experiments.
Firstly, we have identified a regime of nearly Poissonian statistics
of fractional charges, which can be measured through low cumulants.
Beyond the effectively Poissonian regime, we have shown how to extract
the fractional charges and the geometric phase from the waiting time distribution.
Thus, we potentially open up a new road to characterise the topology
of open quantum systems by means of the detector degree of freedom.
An important future project could aim at generalizing this concept to
multidimensional systems, where more than one quantity is measured. Finally, we have shown at a simple,
easily realizable example, that the fractional effect can be used
to detect the presence of multiple charge islands even when the charge
detector cannot distinguish them directly. This could open up a new
research direction to use topological effects in the detector dynamics
for novel measurement techniques. In future research, this effect
could potentially be extended to cases when the resolution of the
discrete process and the detector resolution are incommensurate.

\begin{acknowledgments}
The author is greatly indepted to Janine Splettstoesser and to Maarten
Wegewijs who provided invaluable advice on many of the here treated
topics, and for many highly stimulating discussions. A thank you for
very helpful and interesting discussions goes also to Thomas~L. Schmidt, Sebastian Diehl,
Bj\"orn Sothmann, and Philipp Stegmann.
\end{acknowledgments}

\bibliographystyle{apsrev4-1}
\bibliography{backaction_and_feedback,fractional_QHE,FCS,josephson_junctions,pumping_and_geometric_phase,quantum_thermo,topology_equilibrium,topology_in_transport,topology_nonequilibrium,topology_quantum_info,braid_theory}

\end{document}